\newif\ifShowKeys
\ShowKeystrue
\ShowKeysfalse

\documentclass[11pt]{article}
	\pdfoutput=1
\usepackage[T1]{fontenc}

\usepackage[usenames,dvipsnames]{xcolor}
\usepackage[setpagesize=false,pagebackref=false, 
linktocpage, bookmarksopen=true, colorlinks=true, 
linkcolor=Maroon,citecolor=Maroon,urlcolor=Maroon]{hyperref}

\usepackage[parsep]{collref}				

\ifShowKeys \usepackage[notcite]{showkeys} \fi

\usepackage{amsmath, amssymb,amsthm}
\usepackage{stackrel}
\numberwithin{equation}{section}
\usepackage{bm,environ,mathrsfs,array,arydshln}
\usepackage{booktabs,float,slashed}
\usepackage{appendix}
\usepackage[mathcal]{euscript}
\usepackage{tensor} 						
\usepackage{mathabx}
\usepackage[vcentermath]{youngtab}

\usepackage{graphicx,epsfig,epic}
\usepackage{tikz} 
\usetikzlibrary{arrows,decorations.pathreplacing,decorations.markings,snakes}
\usetikzlibrary{cd}

\usepackage{framed}						
\definecolor{shadecolor}{rgb}{0.9996078, 0.984314, 0.960784}
\definecolor{TFTitleColor}{RGB}{1,1,1}
\definecolor{TFFrameColor}{RGB}{249	218	181}		
\definecolor{TFFrameColor}{RGB}{230 230 230 }

\usepackage{comment}					

\allowdisplaybreaks

\definecolor{myred}{RGB}{233, 33, 45}

\newcommand{\bs}{\begin{shaded}}
\newcommand{\es}{\end{shaded}\noindent}

\def\ba#1\ea{\begin{align}#1\end{align}}		
\newcommand{\be}{\begin{equation}}
\newcommand{\ee}{\end{equation}}
\newcommand{\bea}{\begin{equation} \begin{aligned}} 
\newcommand{\eea}{\end{aligned} \end{equation}}
\newcommand{\mc}{\mathcal }
\newcommand{\wh}{\widehat}
\newcommand{\wt}{\widetilde}

\newcommand{\mk}{\mathfrak}
\newcommand{\la}{\label}
\newcommand{\eps}{\varepsilon}

\newcommand{\lp}{\notag \\ & }

\DeclareMathOperator{\Tr}{\text{Tr}}
\DeclareMathOperator{\vol}{vol}
\DeclareMathOperator{\PE}{PE}

\newcommand{\cf}{\textit{cf.} }
\newcommand{\ie}{\textit{i.e.} }
\newcommand{\eg}{\textit{e.g.} }

\newcommand{\N}{\mathcal N}

\newcommand{\sql}{\sqrt\l}
\renewcommand{\l}{\lambda}

\newcommand{\oneloop}{{\rm 1-loop}}
\newcommand{\F}{{\mathsf F}}
\newcommand{\T}{{\mathsf T}}
\newcommand{\D}{{\mathsf D}}
\newcommand{\I}{\mathrm{I}}
\newcommand{\bz}{\bm{z}}
\newcommand{\smb}{\scalebox{0.6}{$\Box$}}
\newcommand{\vth}{\vartheta}
\newcommand{\ul}{\underline}
\newcommand{\vp}{\varphi}
\newcommand{\vac}{|0\rangle}
\newcommand{\sfM}{\mathsf{M}}

\newcommand{\sfL}{\mathsf{L}}
\newcommand{\sfR}{\mathsf{R}}

\begin{document}



\begin{titlepage}

\vspace*{15mm}
\begin{center}
{\Large\sc   $\N=4$ SYM line defect Schur index and semiclassical string}

\vspace*{10mm}

{\Large M. Beccaria}

\vspace*{4mm}
	
${}^a$ Universit\`a del Salento, Dipartimento di Matematica e Fisica \textit{Ennio De Giorgi},\\ 
		and INFN - sezione di Lecce, Via Arnesano, I-73100 Lecce, Italy
			\vskip 0.3cm
\vskip 0.2cm {\small E-mail: \texttt{matteo.beccaria@le.infn.it}}
\vspace*{0.8cm}
\end{center}

\begin{abstract}  
The giant graviton expansion of the line defect Schur index in four dimensional $\N=4$ $U(N)$ SYM
was recently proposed  in arXiv:2403.11543 to be captured in the dual string theory by  counting fluctuations states of two half-infinite fundamental strings 
in $AdS_{5}\times S^{5}$
ending on the line defect and D3 brane giant. However, agreement with the gauge theory data for the defect line index at finite $N$ 
required the inclusion of ad hoc extra contributions with unclear origin.
We discuss the large $N$ leading order contribution of the giant graviton expansion of the defect line index by a direct analysis of 
semiclassical string partition function in a twisted background. 
We discuss supersymmetric boundary conditions in the presence of the D3 brane and evaluate the quadratic fluctuations 
 effective action by introducing a suitable projection of fluctuation modes. We show that the 
 extra contributions to the single giant graviton correction found in arXiv:2403.11543 correspond to a supersymmetric Casimir energy contribution.
\end{abstract}
\vskip 0.5cm
	{
	}
\end{titlepage}

\tableofcontents
\vspace{1cm}

\section{Introduction}

The superconformal index was introduced in \cite{Kinney:2005ej,Romelsberger:2005eg,Romelsberger:2007ec}  to 
encode all group theoretical information about protected short 
multiplets of  four dimensional  superconformal field theories on $S^{1}\times S^{3}$. Its Schur specialization \cite{Gadde:2011ik,Gadde:2011uv} 
was defined for theories with (at least) $\N=2$ supersymmetry where it computes the vacuum character of the chiral algebra in a 
protected sector \cite{Beem:2013sza}. The Schur index could be generalized in the presence of half-BPS defects, 
in particular line defects \cite{Dimofte:2011py,Gang:2012yr}.

The defect line Schur index is quite interesting from the perspective of AdS/CFT. 
Indeed, in many cases it is exactly computable and thus provides highly non-trivial tests
 of the correspondence  making predictions for string theory in dual geometries associated with defects.

In this paper we consider the Schur index $\I_{\rm F}^{U(N)}$ of 4d $\N=4$ $U(N)$ SYM with two Wilson lines in the 
fundamental and antifundamental of $U(N)$ located at the two poles of $S^{3}$. After conformal mapping 
the two Wilson lines build up a full defect line in  $\mathbb R^{4}$. The line index depends on two fugacities, the universal one $q$ (coupled
to the Hamiltonian and the two spins of the isometry $SO(4)$ of $S^{3}$) and a flavour fugacity $\eta$ coupled to a R-symmetry generator.

The  line index is known exactly at large $N$. In this limit, it factorizes according to 
\be
\la{1.1}
\I_{F}^{U(\infty)} = I_{\rm F1}\, \I_{\rm sugra},
\ee
where $\I_{\rm sugra}$ is the large $N$ limit of the undecorated Schur index. The labels of the two factors in (\ref{1.1})
stem from their AdS/CFT origin. In particular, $\I_{\rm sugra}$ is the index of  KK IIB supergravity modes while $\I_{\rm F1}$ is the Schur index 
of fluctuations of a fundamental string wrapping $AdS_{2}$ in  $AdS_{5}$ and having the Wilson lines as boundary \cite{Gang:2012yr}.

At finite $N$,  superconformal indices admit corrections with $q$-dependent 
weights $\sim q^{N}, q^{2N}, \dots$ multiplied by non-trivial functions of the remaining fugacities. \footnote{
For the purposes of this section our discussion is necessarily schematic.
More precisely, if only $q$ is switched on  the remaining factors are functions of $q$ with a possible 
dependence on $N$ that is at most polynomial. If more fugacities are present, there may be several  expansions
that are non-trivially equivalent \cite{Gaiotto:2021xce,Imamura:2022aua,Fujiwara:2023bdc,Imamura:2024lkw,Imamura:2024fmo}.
} 
In the gauge theory side of AdS/CFT these corrections come from trace relations in the gauge group, 
see \eg \cite{Lee:2023iil}. \footnote{The giant graviton expansion of superconformal indices can be studied on the 
gauge theory by examining finite $N$ dependence without reference to dual string interpretation. It was shown to come from 
counting invariants in generic unitary matrix models \cite{Murthy:2022ien}. It can also be 
interpreted as an instanton expansion in gauge theory  \cite{Eniceicu:2023cxn}. 
The expansion obtained by this approach with the wrapped D-brane expansion, term by term, but the full sum is equivalent, 
see  \cite{Eniceicu:2023uvd}.
}
On the gravity 
side they are due to BPS multiply wrapped branes with charge of order $N$ \cite{Imamura:2021ytr,Gaiotto:2021xce,Lee:2022vig}
and usually called giant gravitons being a generalization of the configurations  studied in the past in 
\cite{McGreevy:2000cw,Mikhailov:2000ya}. In the regime  $N\gg 1$
giant graviton corrections are  exponentially suppressed. 

In the specific case of the Schur index of 4d $\N=4$ $U(N)$ SYM, giant graviton corrections are due to D3 branes wrapped on 
contractible cycles in $S^{5}$ \cite{Arai:2020qaj} with perfect agreement with gauge theory calculations of the index \cite{Beccaria:2024szi}.
For the defect line index, it was recently proposed in \cite{Imamura:2024lkw} that giant graviton corrections come from 
fluctuations of two half fundamental strings wrapping $AdS_{2}$ in  $AdS_{5}$ and ending on the Wilson lines, as well as on 
the D3 giant. This description was shown to be fully consistent with gauge theory computations 
at leading order in the giant graviton expansion \cite{Beccaria:2024oif}. 

However, the gravity calculation in \cite{Imamura:2024lkw} raised a puzzling issue. It 
was based on counting BPS states of fundamental string fluctuations in the 
presence of the D3 brane and required an extra factor 
equal to $1/(\eta q)$ at leading order and whose origin was unexplained.  Schematically, the leading order correction 
had the following form 
\be
\la{1.2}
\delta \I_{F}^{U(N)}(\eta; q) = (\text{D3 fluctuations})\times \frac{1}{\eta q}\times (\text{F1 fluctuations})\, q^{N}+\cdots,
\ee
where the two factors in round bracket represent the naive expected result.
Similar ``extra'' factors were shown to be needed in higher order terms in the 
giant graviton expansion, as well as in more general Schur correlators  \cite{Imamura:2024zvw}.
 
The aim of this paper is to derive (\ref{1.2}) by exploiting the known representation of the superconformal 
index as a supersymmetric partition function, see  \eg  \cite{Nawata:2011un,Aharony:2021zkr}. 
On gauge theory side, this amounts to  study the  Euclidean partition function of the theory on $S^{1}\times S^{3}$ 
with suitable background fields 
associated with the chemical potentials  in the index and preserving supersymmetry 
\cite{Festuccia:2011ws,Dumitrescu:2012ha,Closset:2013vra}. \footnote{The very definition of the index
involves Cartan charges that implement a Scherk-Schwarz dimensional reduction when 4d 
$\N=4$ SYM is obtained from 10d $\N=1$ SYM. }
On gravity side, the same construction requires a corresponding twisted 
deformation of the $AdS_{5}\times S^{5}$ background. In this way, 
the large $N$ giant graviton correction to the superconformal index 
may be computed by a direct semiclassical calculation starting from the Green-Schwarz superstring 
action in static gauge moving in a suitable twisted background that accounts for the 
 charges in the index. 

This approach was recently successfully applied to derive the leading large $N$ correction to several indices. In particular: (a)  the superconformal index of the 6d $(2,0)$ theory on
$S^{1}\times S^{5}$ from quantum M2 brane wrapped on $S^{1}\times S^{2}$ in the dual M-theory twisted background 
$AdS_{7}\times S^{4}$ \cite{Beccaria:2023sph},
(b) the superconformal index of the $\mc N=8$ supersymmetric (level-one) $U(N)\times U(N)$ ABJM theory
from the quantum M5 brane wrapped on $S^{1}\times S^{5}$ in twisted $AdS_{4}\times S^{7}$ background \cite{Beccaria:2023cuo}, and 
(c) the Schur index of 4d $\N=4$ $U(N)$ SYM from quantum D3 brane wrapped on $S^{1}\times S^{3}$ in  twisted 
$AdS_{5}\times S^{5}$ background \cite{Beccaria:2024vfx}.

The case of the defect line index is, however, substantially more difficult than the above examples. In fact, the large $N$ limit of the undecorated Schur index
computes the spectrum of supergravity states, while insertion of defects is  non-trivial on string side, even at large $N$, 
as  illustrated by the
factor $\I_{\rm F1}$ in  (\ref{1.1}). Thus, although   giant graviton corrections to the undecorated index are captured by an extended object
in supergravity background, the same correction in the defect index requires consideration of a more complex system -- here 
the fundamental string(s) and the D3 giant. This requires to clarify novel issues related to the boundary conditions of the 
half-infinite fundamental strings  on the D3 giant,  a key point in the state counting analysis in  \cite{Imamura:2024lkw}. 
We will see that, in our approach, this amounts to evaluating the effective action (or functional determinants) 
of quadratic fluctuations restricted by certain parity conditions or mode selection rules. \footnote{
The techniques we develop are expected to be applicable to more general situations such as 
those recently considered in \cite{Imamura:2024pgp,Imamura:2024zvw}.}

The outcome of our analysis is the explicit semiclassical partition function from quadratic 
bosonic and fermionic fluctuation modes $(X,\theta)$ of the two fundamental half-strings 
in the presence of the spectator D3 giant, \ie
\be
Z = \int DX D\theta \ e^{-S[X,\theta]} = e^{-\Gamma_{\oneloop}^{\rm D3}},
\ee
where the free energy $\Gamma_{\oneloop}^{\rm D3}$ is a function of the flavour fugacity $\eta$ and 
inverse temperature $\beta$, \ie the length of Euclidean thermal cycle with 
$q=e^{-\beta}$. The expression of  $\Gamma_{\oneloop}$  contains two terms
\be
\la{1.4}
\Gamma_{\oneloop}(\eta; \beta)  = \beta\, E_{c}-\sum_{n=1}^{\infty}\frac{1}{n}f_{\rm F}(\eta^{n}, q^{n}),\qquad
E_{c} = -1+\frac{1}{\beta}\log\eta,
\ee
where  $E_{c}$ is the supersymmetric Casimir energy contribution dominating at small temperature while the second term is the 
usual many particle contribution built with a  single particle index $f_{\rm F}(\eta; q)$ that agrees with state  counting in \cite{Imamura:2024lkw}.
We thus  observe that $Z$ 
reproduces (\ref{1.2}) and explains the so far unclear origin of the prefactor $1/(\eta q)$ in terms of a Casimir contribution.
Notice that in the case of the line index
at large $N$, so without D3 giant, the Casimir energy vanishes, consistently with (\ref{1.1}) and absence of extra corrections.
The non vanishing value of $E_{c}$ in the finite $N$ corrections to the line index appears thus to be 
related to the partial breaking of supersymmetry in the presence of the D3 brane.

The fact that $E_{c}$ plays an important role in this matching is peculiar and deserves some comments. 
At the level of gauge theory, 
the relation between the twisted supersymmetric partition function on $S^{1}\times S^{3}$ and the index was shown in many cases to be
\footnote{
Relation (\ref{1.5}) was first observed in \cite{Kim:2012ava} where $E_{c}$ was recognized as the 
supersymmetric version of the Casimir vacuum energy. The possibility of a non-trivial factor connecting the partition function and the index
due to possible local counterterms was discussed in general in \cite{Closset:2013vra}. It was computed explicitly in  \cite{Assel:2014paa} for 
backgrounds with $S^1\times S^{3}$ topology  admitting two supercharges of opposite R-charge, \ie 
Hopf surfaces with two complex structure moduli,
and proved to be scheme independent, as well as independent on continuous parameters in the action in
\cite{Assel:2015nca}. Indeed 
it was related to the anomaly polynomial  in even dimension on $S^{1}\times S^{D-1}$
in \cite{Bobev:2015kza}. For the generalization of $E_{c}$ in 
the presence of 2d conformal defects see  \cite{Chalabi:2020iie,Meneghelli:2022gps}.
}
\be
\la{1.5}
Z(\beta, \mu) = e^{-\beta E_{c}(\mu)}\ \I(\beta\mu),
\ee
where $\mu$ denotes  chemical potentials. The logarithm of the index is exponentially suppressed at large $\beta$ 
and (\ref{1.5}) corresponds to the structure in (\ref{1.4}) and the index is obtained by removing the 
prefactor $e^{-\beta E_{c}}$ from $Z$. On the other hand, the correction (\ref{1.2})
is not an index but instead  its leading giant graviton correction. Besides, as we already remarked, it corresponds to a non trivial
subleading 
semiclassical saddle in the presence of the D3 brane, here  playing an external role. Comparison with (\ref{1.2}) shows 
that the full expression of $Z$ should be kept, including the $E_{c}$ contribution. This is similar to  what happens in 
the Wilson loop calculation in \cite{Beccaria:2023sph}. \footnote{
In \cite{Beccaria:2023sph}, the large $N$ expectation value of a suitable supersymmetric Wilson loop in the SYM theory on $S^{5}$
given by $\langle W \rangle = (2\sinh\frac{\beta}{2})^{-1}e^{N\beta}+\cdots$ was captured by a quantum M2 brane wrapping 
$AdS_{3}$. The factor $e^{N\beta}$ comes from its classical action, while the $\beta$-dependent prefactor is from fluctuations.
In that case, one had, \cf (\ref{1.4}),  $\log(2\sinh\frac{\beta}{2}) = \frac{\beta}{2}-\sum_{n\ge 1}\frac{1}{n}e^{-\beta n}$ where the term $\frac{\beta}{2}$
is the supersymmetric Casimir contribution and it has to be kept.
}
A full clarification of this issue 
presumably needs a non-leading order analysis and is beyond the scope of this paper. 

Our analysis suggests that the study of  finite $N$ giant graviton-like corrections
to defect indices by semiclassical string theory is an interesting approach, likely to be important 
in order to  complement state counting
methods
and capture subtle effects like the middle factor in (\ref{1.2}).

\paragraph{Plan of the paper} In section \ref{sec:schur-intro}  we summarize the available information about the large $N$ limit and leading giant graviton correction to the Schur
index and to the defect line index.
In section \ref{sec:GKL} we discuss general aspects of the computation of superconformal indices and briefly discuss the evaluation of the 
large $N$ line index by state counting.
In section \ref{sec:line-basic} we present in full details its semiclassical evaluation in string theory.
Section \ref{sec:line-giant} considers the leading giant graviton correction to the line index and the novel features associated with 
the boundary conditions of fundamental strings ending on the D3 giant. A few appendices contain  technical material. In particular, in  
Appendix \ref{app:osp} we analyze in some details the ultra-short multiplet of $OSp(4^{*}|4)$ and the structure of the small superconformal representations
in the presence of the D3 brane.

\section{Schur (line) index and leading giant graviton correction}
\la{sec:schur-intro}

The definition of the Schur index in 4d $\N=4$ $U(N)$ SYM is
\be
\la{2.1}
\I^{U(N)}(\eta; q) = \Tr_{\rm BPS}[(-1)^{\rm F}\, q^{H+J+\bar J}\eta^{R}],
\ee
where  $H$ is the Hamiltonian, $J, \bar J$ are two spins, \footnote{The theory in $\mathbb R^{4}$ is radially quantized in $\mathbb{R}
\times S^{3}$ and $J, \bar J$ are Cartan of $SO(4)\simeq SU(2)\times SU(2)$ 
where $SO(4)$ is the isometry group of the spatial part $S^{3}$.} 
and  $R$ is a generator of the $SU(4)_{R}$ R-symmetry of the $PSU(2,2|4)$ superconformal group.
The variables $q, \eta$ are the so-called universal and flavor fugacities.
The Schur index admits the  holonomy integral representation \cite{Gaiotto:2019jvo} \footnote{
The measure is $D^{N}\bz = \frac{1}{N!}\prod_{n=1}^{N}
\frac{dz_{n}}{2\pi i\, z_{n}}\prod_{n\neq m}(1-\frac{z_{n}}{z_{m}})$. The character $\chi_{\smb}(\bz)$ of the $U(N)$ fundamental representation is $\chi_{\smb}(\bz) = \sum_{n=1}^{N}z_{n}$. The operation $\PE$ is plethystic exponentiation with respect to fugacities
and holonomies.}
\ba
\la{2.2}
\I^{U(N)}(\eta; q) & = {\oint}_{|\bz|=1}D^{N}\bz\, \PE[ f(\eta; q)\chi_{\smb}(\bz)\chi_{\smb}(\bz^{-1})],\qquad
f(\eta; q) = \frac{(\eta+\eta^{-1})\, q-2q^{2}}{1-q^{2}},
\ea
where the function $f(\eta; q)$ is the single particle Schur index.
Exact results for the Schur index were obtained  in the case of $U(N)$ gauge group in \cite{Bourdier:2015sga,Bourdier:2015wda,Pan:2021mrw} for $\eta=0$ and in \cite{Hatsuda:2022xdv} for $\eta\neq 0$, 
and generalized to $B_{n}$,$C_{n}$, $D_{n}$, $G_{2}$  groups in  \cite{Du:2023kfu}.
The leading large $N$ correction to the Schur index reads \footnote{Notation for $q$-functions is 
$(a; q)_{\infty}= \prod_{k=0}^{\infty}(1-a\,q^{k})$, $(a^{\pm}; q)_{\infty} = (a; q)_{\infty}(a^{-1};  q)_{\infty}$, 
$(q)_{\infty} \equiv (q; q)_{\infty} = \prod_{k=1}^{\infty}(1-q^{k})$, and 
$\vth(x,q) = -x^{-\frac{1}{2}}(q)_{\infty}(x; q)_{\infty}(qx^{-1}; q)_{\infty}$.
}
\be
\la{2.3}
\frac{\I^{U(N)}(\eta; q)}{\I^{U(\infty)}(\eta; q)} = 
1 +\bigg[\eta^{N}G^{+}_{\rm D3}(\eta; q)+\eta^{-N}G^{-}_{\rm D3}(\eta; q)\bigg]\, q^{N}+\mc O(q^{2N}),
\quad \I^{U(\infty)}(\eta; q) = \frac{(q^{2})_{\infty}}{(\eta q)_{\infty}(\eta^{-1}q)_{\infty}},
\ee
where the function $G^{\pm}_{\rm D3}(\eta; q)$ is the single giant graviton contribution from wrapped D3 brane \cite{Arai:2020qaj,Beccaria:2024szi} and admits the closed expression 
\be
G^{+}_{\rm D3}(\eta; q) = G^{-}_{\rm D3}(\eta^{-1}; q)  = -\eta^{2}q\, \frac{(\frac{q}{\eta})_{\infty}^{3}}{\vth(\eta^{2},\frac{q}{\eta})}.
\ee
As we discussed in the Introduction, an interesting generalization of the Schur index (\ref{2.1}) 
consists in decorating it by inserting BPS defect lines \cite{Dimofte:2011py,Gang:2012yr} and  exact results were obtained in the presence of 
an arbitrary number of 't Hooft or Wilson lines in \cite{Drukker:2015spa,Cordova:2016uwk,Neitzke:2017cxz,
Hatsuda:2023iwi,Hatsuda:2023imp,Hatsuda:2023iof,Guo:2023mkn}.  
\footnote{The expression corresponding to (\ref{2.2})
with Wilson line defects in representations $R_{1}, R_{2}, \dots$ is 
$\I^{U(N)}_{R_{1}, R_{2}, \dots}(\eta; q)  = {\oint}_{|\bz|=1}D^{N}\bz\, \prod_{n\ge 1}\chi_{R_{n}}(\bz)\, 
\PE[ f(\eta; q)\chi_{\smb}(z)\chi_{\smb}(z^{-1})]$.}
The associated so-called 
Schur (line defect) correlators may be regarded as a supersymmetric partition function on $S^{1}\times S^{3}$ with  defect lines 
wrapping $S^{1}$ and located on a great circle of $S^{3}$ to preserve the same supersymmetry used in definition 
of the undecorated index \cite{Cordova:2016uwk}. Schur 
correlators are topological and do not depend on the precise position of the insertions (at fixed ordering of insertions).
We will focus on the line defect 2-point function $\I^{U(N)}_{\rm F}(\eta; q)\equiv \I^{U(N)}_{\smb, \overline{\smb}}(\eta; q)$
with two Wilson lines one in the fundamental and the other in the anti-fundamental.
In the large $N$ limit it takes the factorized form \cite{Gang:2012yr}
\be
\la{2.5}
\I_{\rm F}^{U(\infty)}(\eta; q) = \I_{\rm F1}(\eta; q)\  \I^{U(\infty)}(\eta; q), \qquad  \I_{\rm F1}(\eta; q) = \frac{1}{1-f(\eta; q)}\,.
\ee
The algebraic identity 
\be
\la{2.6}
\I_{\rm F1}(\eta; q) = \PE[f_{\rm F1}(\eta; q)], \qquad f_{\rm F1}(\eta; q) = -q^{2}+(\eta+\eta^{-1})\, q,
\ee
has a clean AdS/CFT interpretation. The factor $\I_{\rm F1}(\eta; q)$  
is expected to represent fluctuations of a fundamental string along $AdS_{2}\subset AdS_{5}$ \cite{Rey:1998ik,Maldacena:1998im}
 meeting the boundary of $AdS_{2}$ at the two poles of $S^{3}$
in $\partial AdS_{5} = \mathbb{R}\times S^{3}$, where we place the two line operators are placed. 
This was checked in  \cite{Gang:2012yr} where the expression $f_{\rm F1}(\eta; q)$ was shown to perfectly match the single particle index of fluctuations of the fundamental string. 

The leading large $N$ correction to the line index was first addressed in \cite{Imamura:2024lkw}
on string side assuming it is given by fluctuations of two half-infinite strings ending on the D3 giant. At the single giant graviton order, 
it was later confirmed 
on gauge theory side at all orders in $q$ in \cite{Beccaria:2024oif}. 
It takes a form similar to  (\ref{2.3}), and it is convenient to present it as a correction to the ratio
\be
R_{N}(\eta; q) = \frac{\I_{\rm F}^{U(N)}-\I_{\rm F1}\,\I^{U(N)}}{\I^{U(\infty)}} .
\ee
The numerator of this ratio subtracts the contribution to $\I_{\rm F1}\I^{(N)}$  corresponding to the case when the 
two half-infinite strings do not end on the giant graviton and the difference is divided by the supergravity contribution,
\ie the undecorated index $\I^{U(\infty)}$ at large $N$. At leading large $N$, the exact expression of the first correction reads
\bea
\la{2.8}
& R_{N}(\eta; q)   = 
\bigg(\mathscr G^{+}_{\rm F}(\eta; q)\, \eta^{N}+\mathscr G^{-}_{\rm F}(\eta; q)\,\eta^{-N}\bigg)\, q^{N}+\mc O(q^{2N}),
\eea
where the non-trivial correction factors $\mathscr G^{\pm}(\eta; q)$ are given by 
\bea
\la{2.9}
\mathscr G^{\pm}_{\rm F}(\eta; q)  =  G^{\pm}_{\rm D3}(\eta; q)\times \frac{1}{\eta q}\,  \PE[f_{\rm F}(\eta; q) ]\, , 
\qquad f_{\rm F}(\eta; q) = 2\eta^{-1}q-2q^{2}\, .
\eea
As observed in \cite{Imamura:2024lkw} by counting BPS contributions to the index,  
the function   $f_{\rm F}(\eta; q)$  in (\ref{2.9}) agrees with the single particle index from fluctuations of the two semi-infinite strings 
ending on the giant graviton. Factorization in (\ref{2.9}) implies  an important decoupling of the contributions from fluctuations
of the D3 giant, at least at leading order in large $N$.

The extra factor $1/(\eta q)$ in (\ref{2.9}) is a puzzling feature and so far has no explanation. As we mentioned
in the Introduction, it seems important to clarify its origin 
because similar (more complicated) extra factors  are present in  higher order giant graviton contributions to the line index and its
generalizations 
\cite{Imamura:2024lkw,Imamura:2024zvw}.

\section{Review of Gang-Koh-Lee calculation}
\la{sec:GKL}

We begin with a quick summary of the computation in   \cite{Gang:2012yr} and a review of general aspects
concerning the evaluation of  superconformal indices on the two sides of AdS/CFT.
The line index $\I_{\F}(q; \eta)$ in $\N=4$ SYM is defined starting  with a preserved supercharge $Q$ such that
\be
\Delta \equiv \{Q, Q^{\dagger}\} = H-J-\bar J-r_{1}, \qquad [Q, H+J+\bar J]=0,
\ee
where $H$ is the energy associated with time direction in $\mathbb{R}\times S^{3}$ 
or conformal dimension of operators in $\mathbb{R}^{4}$.
The two angular momenta $J, \bar J$ are Cartan of $SU(2)_{L}$, $SU(2)_{R}$. Finally $r_{1,2,3}$ are Cartan of R-symmetry $SU(4)$. 
The line index  (\ref{2.1}) is
\be
\la{3.2}
\I_{\F}(q; \eta) = \Tr_{\mc H_{L}}(-1)^{F}q^{H+J+\bar J}\eta^{r_{3}},
\ee
where $\mc H_{L}$ is the Hilbert space on $S^{3}$ in the presence of line operator $L$ and the trace is restricted to BPS states
with $\Delta=0$.
In the gauge theory side, the actual calculation of the index can be done in two complementary ways.  A direct method consists just in 
summing over BPS states with $\Delta=0$. Alternatively, the index can be regarded as a supersymmetric 
partition function on $S^{1}_{\beta}\times S^{3}$
after Wick rotation $\tau=-it$ and compactification $\tau=\tau+\beta$. Fields are periodic along $S^{1}_{\beta}$ and 
should be expanded in fluctuations around a background suitable for the line operators $L$ and chemical potentials.  In our context,
this corresponds to the following twist in the action 
\be
\la{3.3}
\partial_{\tau}\to\partial_{\tau}-(j_{L}+j_{R})+r_{3}\frac{\log \eta}{\beta}.
\ee
These two approaches have a counterpart on the gravity side. The calculation in  \cite{Gang:2012yr}
is done by the direct method. The gravity dual in the presence of two oppositely charged Wilson lines in the fundamental and anti-fundamental is a fundamental string wrapping $AdS_{2}$ in  $AdS_{5}$ with metric
\be
ds^{2}_{AdS_{5}} = d\rho^{2}+\cosh^{2}\rho\, d\tau^{2}+\sinh^{2}\rho\, dS_{3}.
\ee
The string  worldsheet $(\tau, \rho)$ is at fixed poles of $S^{3}$ $(1,0,0,0)$ and $(-1,0,0,0)$ (and at a point in $S^{5}$). It 
preserves the same  symmetries of a single Wilson line in the CFT theory.
The symmetry group of $\N=4$ SYM is $PSU(2,2|4)$ with bosonic $SU(2,2)\times SU(4)_{R}$. The first factor
is the conformal group in 4d $SU(2,2)\simeq SO(4,2)$. The second is $SU(4)_{R}\simeq SO(6)_{R}$ and gives R-symmetry 
equal to isometries of $S^{5}$.
The symmetry group of the line Wilson loop is discussed in \cite{Gomis:2006sb}. 
The loop breaks part of the $SO(4,2)$ generators $\{P_{\mu}, J_{\mu\nu}, D, K_{\mu}\}$. In fact, 
invariance of the line (along 0 direction) is preserved by $\{P_{0}, J_{ij}, D, K_{0}\}$. The generators $J_{ij}$ span 
$SO(3)\simeq SU(2)$\footnote{Here and in the following we denote by $\simeq$ the standard 2-1 homomorphisms.} 
(rotations around the line).
The other ones give rotation around the line, dilatation, and a special transformation. Generators close as 
\be
[P_{0},K_{0}]=-2D, \qquad [P_{0},D]=-P_{0}, \qquad [K_{0},D]=K_{0},
\ee
and span $SU(1,1)= SL(2, \mathbb R)$. Thus  conformal symmetry is reduced to  
\be
SO(4,2)\to SL(2, \mathbb R)\times SO(3)\simeq SO(4^{*})
\ee
where we recall that $SO(2n^{*})$ is a non-compact real form of $SO(2n,\mathbb C)$.
The R-symmetry group is broken to $SO(6)_{R}\to SO(5)_{R} \simeq USp(4)_{R}$ by the choice of the versor in the coupling of the loop
with $\N=4$ SYM scalars.
Finally the Wilson line is $1/2$-BPS and preserves 16 of the supersymmetries of $PSU(2,2|4)$. The AdS supergroup with 16 supercharges
and bosonic part $SO(4^{*})\times USp(4)_{R}$ is $OSp(4^{*}|4)$ \cite{Nahm:1977tg,Gunaydin:1986fe}.
Fluctuations of the fundamental string were computed in \cite{Drukker:2000ep} and indeed can be arranged into  
8 bosonic and 8 fermionic states in the $OSp(4^{*}|4)$ ultra-short multiplet \cite{Faraggi:2011bb} 
\be
\la{3.7}
(1,\bm{1},\bm{5})+(\tfrac{3}{2},\bm{2},\bm{4})+(2,\bm{3},\bm{1}),
\ee
where the first label $h$ is  eigenvalue of dilatation in  $SL(2,\mathbb R)$, the second label is dimension of $SU(2)$ 
with angular momentum  $\mc J = J+\bar J$, and the third label is the dimension of $USp(4)_{R}$. \footnote{Values of $h$ 
can be checked to agree
with the explicit mass spectrum in \cite{Drukker:2000ep}. For instance, 
for scalars one finds  three massive fluctuations in $AdS_{5}$ with $m^2 = 2$ and five massless 
fluctuations in $S^{5}$ with $m^2 = 0$. For scalars, we use $h = \frac{1}{2} (d + \sqrt{d^2+4m^2})$
where here for $AdS_{1+1}$ we have $d=1$. So $m^2 = 0$ gives $h=1$ and $m^2 = 2$ gives $h =  2$ as in (\ref{3.7}).}

Following the direct counting approach,  the trace (\ref{3.2}) is evaluated in \cite{Gang:2012yr}
by summing over fluctuations modes after restriction to BPS states.
For the states in the $\bm{5}$ of $SO(5)$ one has 
\be
\bm{5}: (r_{1},r_{3}) = (1,1), (1,-1), (-1,-1), (-1,1), (0,0),   \qquad     h = 1, \mc J=0,
\ee
so we have two bosonic BPS states with $(r_{1},r_{3}) = (1,1), (1,-1)$.
For the states in the $\bm{4}$ of $SO(5)$ one has 
\be
\bm{4}: (r_{1},r_{3}) = (1,0), (-1,0), (0,1), (0,-1),  \qquad   h=\tfrac{3}{2}, \mc J=\tfrac{1}{2},
\ee
so we have one fermionic BPS state with $(r_{1},r_{3}) = (1,0)$.
From the singlet of SO(5) we don't have BPS states. In total, the index gets the following three contributions
\bea
(1,0,5) \ \text{with}\  (r_{1},r_{3}) = (1,1), &\qquad\to  (-1)^F q^{h+\mc J} \eta^{r_{3}} = q \eta, \\
(1,0,5) \ \text{with}\ (r_{1},r_{3}) = (1,-1), &\qquad\to (-1)^F q^{h+\mc J} \eta^{r_{3}} = q \eta^{-1}, \\
(\tfrac{3}{2}, \tfrac{1}{2},4)\ \text{with}\  (r_{1},r_{3}) = (1,0), &\qquad\to  (-1)^F q^{h+\mc J} \eta^{r_{3}} = -q^2,
\eea
summig up to the single particle index $f_{\rm F1}(\eta;q)$ in (\ref{2.6}).

In the following sections, we will begin by obtaining this result by the indirect approach (still on gravity side), \ie by 
evaluating the string  semiclassical  partition function in a suitable background. Later, we will address finite $N$ corrections
in the same framework, but in the presence of the  D3 giant.

\section{Large $N$ line index from $AdS_{2}$ string in twisted background}
\la{sec:line-basic}

Let us introduce the following  twists in the $AdS_{5}\times S^{5}$ background associated with the Cartan generators in the index. 
A first twist takes into account 
$\mc J = J+\bar J$ and is a rotation of an angle in $S^{3}\subset AdS_{5}$. A second one is for $r_{3}$ in the index
and amounts to two opposite rotations of two angles in $S^{5}$ in toroidal parametrization, \cf Section 5 of \cite{Beccaria:2024vfx}. The twisted background is thus
\ba
ds^{2}_{\wt{AdS}_{5}} &= d\rho^{2}+\cosh^{2}\rho\, d\tau^{2}+\sinh^{2}\rho\, d\wt{S}_{3}, \\
d\wt{S}_{3} &= d\psi_{1}^{2}+\sin^{2}\psi_{1}d\psi_{2}^{2}+\sin^{2}\psi_{1}\sin^{2}\psi_{2}(d\psi_{3}+i\alpha_{1}d\tau)^{2}, \\
\la{4.3}
ds^{2}_{\wt{S}^{5}} &= 
dn_{1}^{2}+n_{1}^{2}d\vp_{1}^{2}+
dn_{2}^{2}+n_{2}^{2}(d\vp_{2}-i\alpha_{2} d\tau)^{2}+
dn_{3}^{2}+n_{3}^{2}(d\vp_{3}+i\alpha_{2} d\tau)^{2},\\
& n_{1}^{2}+n_{2}^{2}+n_{3}^{2}=1. \notag
\ea
The specific value of the twist angles $\alpha_{1}, \alpha_{2}$ will be discussed later.
We take as classical solution the N pole or S pole in $S^{3}$ so $\psi_{1}=0,\pi$, with both giving the same.
The choice of a specific point in $S^{5}$ is  irrelevant by rotational symmetry.

The fundamental string is wrapped on $AdS_{2,\beta}$ with coordinates
 $(\xi^{1},\xi^{2}) = (\tau\equiv \tau+\beta,\rho)$. The bosonic action is 
\ba
S_{\rm NG} &= \T\int d^{2}\xi\sqrt{G}, \qquad G_{ab}=\partial_{a}X^{M}\partial_{b}X^{N}G_{MN}(X) = 
G_{ab}(\xi)\,d\xi^{a}d\xi^{b},
\ea
where $\T=\frac{\sql}{2\pi}\gg 1$ controls the semiclassical expansion. 
Given a classical solution $X^{M}=X^{M}(\xi)$ ($M=1, \dots, 10$) we adopt a static gauge where two of the $X^{M}$ 
coordinates are equal to the  world-volume coordinates $\xi^{a}$ ($a=1,2$) and  a $\kappa$-symmetry gauge for  fermions. 
The remaining 8 bosonic and 8 fermionic fluctuations  produce a $\beta$-dependent 1-loop prefactor in the  partition function $\rm Z$
\ba
{\rm Z} &= \int DXD\theta\ e^{-S[X,\theta]} = \mc Z_{1}\, e^{-\T\, \bar{S}_{\rm cl}}[1+\mc O(\T^{-1})], \qquad S_{\rm cl} = \T\bar{S}_{\rm cl}, \\
\mc Z_{\oneloop}&= e^{-\Gamma_{\oneloop}}, \qquad \Gamma_{\oneloop} = \frac{1}{2}\sum_{k}(-1)^{{\F}_{k}}\log\det\Delta_{k},
\ea
where $\Delta_{k}$ is the  differential operator governing quadratic fluctuations of the $k$-th field. 
In our case, the classical action vanishes because the Wilson line is BPS \cite{Drukker:2000ep}. \footnote{With the usual 
regularization of the boundary, this follows from
$\int_{0}^{-\log \eps}\cosh\rho\, d\rho = \tfrac{1}{2\eps}+0+\mc O(\eps)$.
} The aim is thus to compute the non-trivial piece $\Gamma_{\oneloop}=\Gamma_{\oneloop}(\eta; \beta)$.

\subsection{Scalar fluctuations}

\subsubsection{$AdS_{5}$ sector}

To study bosonic fluctuations in $AdS_{5}$ 
we introduce new variables to parametrize $S^{3}\subset AdS_{5}$
\begin{alignat}{2}
& X_{1} =  \cos\psi_{1}, && X_{2} =  \sin\psi_{1}\, \cos\psi_{2}, \\
& X_{3} =  \sin\psi_{1}\, \sin\psi_{2}\, \cos\psi_{3}, \ \ \ \ &&  X_{4} =  \sin\psi_{1}\, \sin\psi_{2}\, \sin\psi_{3},
\end{alignat}
with $X_{1}^{2}+\cdots+ X_{4}^{2}=1$. We have at quadratic order ($I=2,3,4$)
\be
ds^{2}_{\wt{AdS}_{5}} = d\rho^{2}+\cosh^{2}\rho \, d\tau^{2}+\sinh^{2}\rho \, [dX_{I}^{2}-\alpha_{1}^{2}(X_{3}^{2}+X_{4}^{2})d\tau^{2}
+2i\alpha_{1}(X_{3}dX_{4}-X_{4}dX_{3})d\tau].
\ee
Denoting by $a=1,2$ the indices of the world-sheet coordinates $\xi^{a}=(\tau,\rho)$, we obtain 
\ba
ds^{2}_{\wt{AdS}_{5}} &= d\rho^{2}+[\cosh^{2}\rho-\alpha_{1}^{2}\sinh^{2}\rho(X_{3}^{2}+X_{4}^{2})+2i\alpha_{1}\sinh^{2}\rho(X_{3}\dot X_{4}-X_{4}\dot X_{3})] \, d\tau^{2}
\lp
+\sinh^{2}\rho\, \partial_{a}X_{I}\partial_{b}X_{I}\, d\xi^{a}d\xi^{b}
+\text{off diagonal terms}\,.
\ea
Expanding the metric by 
$\delta\sqrt{G^{(0)}+G^{(1)}} = \sqrt{G^{(0)}}+\frac{1}{2}\sqrt{G^{(0)}}\, (G^{(0)})^{ab}G^{(1)}_{ab}+\cdots$,
we get 
\be
\delta S_{\rm NG} = \frac{1}{2}\T\int d^{2}\xi\sqrt{g}[g^{ab}\, \sinh^{2}\rho\,
\partial_{a}X_{I}\partial_{b}X_{I}-\alpha_{1}^{2}\tanh^{2}\rho\,(X_{3}^{2}+X_{4}^{2})+2i\alpha_{1}\tanh^{2}\rho\, (X_{3}\dot X_{4}-X_{4}\dot X_{3})],
\ee
where $g_{ab}\equiv G^{(0)}_{ab}$ is the $AdS_{2}$ metric 
\be
\la{4.14}
ds^{2}_{AdS_{2}} = g_{ab}d\xi^{a}d\xi^{b} = d\rho^{2}+\cosh^{2}\rho\, d\tau^{2}, \qquad g_{ab}=\begin{pmatrix} 1 & 0 \\ 0 & \cosh^{2}\rho\end{pmatrix},
\qquad \sqrt g = \cosh\rho.
\ee
Notice that for small $\rho$ one has $ds^{2}\simeq d\rho^{2}+d\tau^{2}$ or $S^{1}\times S^{1}_{\beta}$ and the thermal cycle 
is not contractible.
With the field redefinition
\be
X_{I}= \frac{1}{\sinh\rho} Y_{I},
\ee
we get
\ba
g^{ab}\, \sinh^{2}\rho\,
\partial_{a}X_{I}\partial_{b}X_{I} = \sinh^{2}\rho\partial_{\rho}\bigg(\frac{Y_{I}}{\sinh\rho}\bigg)\partial_{\rho}\bigg(\frac{Y_{I}}{\sinh\rho}\bigg)+
\frac{1}{\cosh^{2}\rho}\dot Y_{I}\dot Y_{I},
\ea
and integration by parts leads to 
\be
\delta S_{\rm NG} = \frac{1}{2}\beta\T\int d^{2}\xi\sqrt g[g^{ab}\partial_{a}Y_{I}\partial_{b}Y_{I}+2\,Y_{I}^{2}
-\frac{\alpha_{1}^{2}}{\cosh^{2}\rho}(Y_{3}^{2}+Y_{4}^{2})+2i\frac{\alpha_{1}}{\cosh^{2}\rho}(Y_{3}\dot Y_{4}-Y_{4}\dot Y_{3})].
\ee
The twist $\alpha_{1}$ gives  $\rho$ dependent mass terms (and mixing contributions)  for two of the three scalars. This is same as 
coupling the complex scalar $Y_{3}+i Y_{4}$  to a constant gauge field in $\tau$ direction. Indeed, let us define
\be
Z = \frac{1}{\sqrt 2}(Y_{3}+iY_{4}).
\ee
We can write
\ba
\dot Y_{3}\dot Y_{3}+\dot Y_{4}\dot Y_{4}+Y_{3}^{2}+Y_{4}^{2}-\alpha_{1}^{2}(Y_{3}^{2}+Y_{4}^{2})+2i\alpha_{1}(Y_{3}\dot Y_{4}-Y_{4}\dot Y_{3}) = 
2(\dot{\bar Z}+\alpha_{1}\bar Z) (\dot Z-\alpha_{1} Z)+4\bar Z Z,
\ea
and thus
\be
\delta S_{\rm NG} = \T\int d^{2}\xi\sqrt g\, \bigg[\frac{1}{2}\bigg(g^{ab}\partial_{a}Y_{2}\partial_{b}Y_{2}+2Y_{2}^{2}\bigg)
+g^{ab}D_{a}\bar Z D_{b}Z+2\bar Z Z\bigg],
\ee
where $D_{a}Z = \partial_{a}Z - A_{a}Z$, $D_{a}\bar Z = \partial_{a}\bar Z + A_{a}\bar Z$ with the constant gauge field along $\tau$
\be
A_{a} = (\alpha_{1},0).
\ee
The squared mass in $AdS_{2}$ is same $m^{2}=2$ for all three ($1+2$) fluctuations. The $\alpha_{1}$ twist enters only
through the constant gauge field.

\subsubsection{$S^{5}$ sector}

In unflavored case $\eta=1$ we  switch-off the twist in $S^{5}$ and get five massless scalars in (thermal) $AdS_{2}$ as follows from \cite{Drukker:2000ep}.
In flavored case, we replace in (\ref{4.3}) the explicit parametrization
%
\bea
&(n_{1}\sin\vp_{1}, n_{1}\cos\vp_{1}) = (U_{1},U_{2}), \\
&(n_{2}\sin\vp_{2}, n_{2}\cos\vp_{2}) = (U_{3},U_{4}), \\
&(n_{3}\sin\vp_{3}, n_{3}\cos\vp_{3}) = (U_{5},U_{6}), \qquad U_{1}^{2}+\cdots +U_{6}^{2}=1, \\
& \vp_{1}=\arctan\frac{U_{1}}{U_{2}}, \qquad
 \vp_{2}=\arctan\frac{U_{3}}{U_{4}}, \qquad
 \vp_{3}=\arctan\frac{U_{5}}{U_{6}}, \\
n_{1} &= \sqrt{U_{1}^{2}+U_{2}^{2}}, \qquad
n_{2} = \sqrt{U_{3}^{2}+U_{4}^{2}}, \qquad
n_{3} =  \sqrt{U_{5}^{2}+U_{6}^{2}}.
\eea
Solving $U_{1}$ in terms of the other fluctuation fields $U_{2}, \dots, U_{6}$, we obtain 
\ba
ds^{2}_{\wt{S}^{5}} &= \sum_{I=2}^{6} dU_{I}^{2}+2i\alpha_{2}(U_{3}dU_{4}-U_{4}dU_{3}-U_{5}dU_{6}+U_{6}dU_{5})\, d\tau
-\alpha_{2}^{2}(U_{3}^{2}+U_{4}^{2}+U_{5}^{2}+U_{6}^{2})\,d\tau^{2}.
\ea
Introducing two complex scalars
\be
W_{+} = \frac{U_{3}+iU_{4}}{\sqrt 2}, \qquad
W_{-} = \frac{U_{5}+iU_{6}}{\sqrt 2}, 
\ee
gives (following similar steps as in AdS sector)
\be
\delta S_{\rm NG} = \T\int d^{2}\xi\sqrt g\, \bigg[\frac{1}{2}\bigg(g^{ab}\partial_{a}U_{2}\partial_{b}U_{2}\bigg)
+\sum_{s=\pm} g^{ab}D_{a}\overline W_{s} D_{b}W_{s}\bigg],
\ee
where the covariant derivative couples $W_{\pm}$ to $\pm A'_{a}$ which is 
the following constant gauge field along Euclidean time direction $\tau$
\be
A'_{a} = (\alpha_{2},0).
\ee
In summary, in the sector of scalar fluctuations we have a massless scalar plus two complex scalars coupling to $A'$ with  opposite charge.

\subsection{Fermionic fluctuations}

We have 8 fermionic fields with conformal dimension $h=\frac{3}{2}$ in the ultra-short $OSp(4^{*}|4)$ representation.
The detailed coupling to a constant gauge field
can be derived from the fermionic part of the Green-Schwarz action \cite{Metsaev:1998it}, see also \cite{Forini:2015mca}.

\paragraph{Fermionic action} 

The fermionic action is ($\bar\theta = \theta^{\rm T}C$) \footnote{
There were several  factors of 2
missing or notation not fully explained in some earlier papers
but end results were  correct, see footnote 34 in revised version of  \cite{Arutyunov:2015mqj}.
}
\be
L_{F} = i\,(\sqrt g\, g^{ab}\,\delta^{IJ}-\eps^{ab}s^{IJ})\,\bar\theta^{I}\rho_{a}D_{b}^{JK}\theta^{K},
\qquad \rho_{a}=\Gamma_{\ul{m}}e^{\ul{m}}_{a}, \qquad e^{\ul{m}}_{a}=E^{\ul{m}}_{\mu}\partial_{a}X^{\mu},
\ee
where $I,J=1,2$, $s^{IJ}=\text{diag}(1, -1)$, and $\rho_{a}$ are projections of the 10-d Dirac matrices. The fermionic fields
are two 10d MW spinors $\theta^{I}$ with same chirality.  $X^{\mu}$ are the string coordinates, 
\ie given functions of $\tau$ and $\sigma$ for a particular classical solution.

The covariant derivative takes the following form 
\bea
D_{a}^{IJ} &= \delta^{IJ}\D_{a}+\mc F_{a}\eps^{IJ}, \\
\D_{a}&=\partial_{a}+\tfrac{1}{4}\omega^{\ul{mn}}_{a}\Gamma_{\ul{mn}},\qquad \omega_{a}^{\ul{mn}}=\partial_{a}X^{\mu}\omega_{\mu}^{\ul{mn}},
\eea
where the flux term is 
\be
\mc F_{\mu} = -\frac{1}{8\cdot 5!}F_{\mu_{1}\dots\mu_{5}}\Gamma^{\mu_{1}\dots\mu_{5}}\Gamma_{\mu}.
\ee
We fix $\kappa$-symmetry by 
\be
\theta^{1}=\theta^{2}=\theta.
\ee
Then
\bea
L_{F} &=L_{F}^{\rm kin}+L_{F}^{\rm flux}, \\
L_{F}^{\rm kin} &= 2i\,\sqrt g g^{ab}\bar\theta \rho_{a}(\partial_{b}+\tfrac{1}{4}\omega^{\ul{mn}}_{b}
\Gamma_{\ul{mn}})\theta, \qquad
L_{F}^{\rm flux} = -2i\eps^{ab}\bar\theta \rho_{a}\mc F_{b}\theta.
\eea

\paragraph{Specialization to our background}

In real time $\tau = it$
\bea
ds^{2}_{\wt{AdS}_{5}} &= d\rho^{2}-\cosh^{2}\rho\, dt^{2}+\sinh^{2}\rho\, d\wt{S}_{3}, \\
d\wt{S}_{3} &= d\psi_{1}^{2}+\sin^{2}\psi_{1}d\psi_{2}^{2}+\sin^{2}\psi_{1}\sin^{2}\psi_{2}(d\psi_{3}-\alpha_{1}dt)^{2}, \\
ds^{2}_{\wt{S}^{5}} &= 
dn_{1}^{2}+n_{1}^{2}d\vp_{1}^{2}+
dn_{2}^{2}+n_{2}^{2}(d\vp_{2}+\alpha_{2} dt)^{2}+
dn_{3}^{2}+n_{3}^{2}(d\vp_{3}-\alpha_{2} dt)^{2}.
\eea
In classical solution parametrized by $(t,\rho)$ all other coordinates are zero with the exception of $n_{1}=1$. 
Let us introduce angles for the $\bm{n}$ 3-vector
\be
n_{1}=\cos\chi_{1}, \qquad n_{2} = \sin\chi_{1}\cos\chi_{2}, \qquad n_{3}=\sin\chi_{1}\sin\chi_{2},
\ee
The metric is 
\ba
ds^{2}_{\wt{AdS}_{5}} &= -\cosh^{2}\rho\, dt^{2}+d\rho^{2}+\sinh^{2}\rho\, d\wt{S}_{3}, \nonumber \\
d\wt{S}_{3} &= d\psi_{1}^{2}+\sin^{2}\psi_{1}d\psi_{2}^{2}+\sin^{2}\psi_{1}\sin^{2}\psi_{2}(d\psi_{3}-\alpha_{1}dt)^{2}, \\
ds^{2}_{\wt{S}^{5}} &= 
d\chi_{1}^{2}+\sin^{2}\chi_{1}d\chi_{2}^{2}+\cos^{2}\chi_{1}d\vp_{1}^{2}
+\sin^{2}\chi_{1}\cos^{2}\chi_{2}(d\vp_{2}+\alpha_{2}dt)^{2}+
\sin^{2}\chi_{1}\sin^{2}\chi_{2}(d\vp_{3}-\alpha_{2}dt)^{2}.\nonumber
\ea
Let us label 10d coordinates as 
\be
\def\arraystretch{1.3}
\begin{array}{ccccc|ccccc}
\toprule
t & \rho & \psi_{1} & \psi_{2} & \psi_{3} & \chi_{1} & \chi_{2} & \vp_{1} & \vp_{2} & \vp_{3} \\
0 & 1 & 2 & 3 & 4  & 5 & 6 & 7 & 8 & 9 \\
\bottomrule
\end{array}\notag
\ee
Vielbein $E^{\ul{m}} = E^{\ul{m}}_{\mu}dX^{\mu}$ (with flat 10d metric in $(-,+,+, \dots, +)$ signature) are  
\bea
& E^{\ul 0} = \cosh\rho\,dt \ \ E^{\ul 1} = d\rho, \ \ E^{\ul 2}=\sinh\rho d\psi_{1}, \ \ E^{\ul 3} = \sinh\rho\sin\psi_{1} d\psi_{2}, 
\ \ E^{\ul 4} =  \sinh\rho\sin\psi_{1}\sin\psi_{2} (d\psi_{3}-\alpha_{1}dt), \\
& E^{\ul 5} = d\chi_{1}, \ \ E^{\ul 6} = \sin\chi_{1}d\chi_{2}, \ \ E^{\ul 7} = \cos\chi_{1}d\vp_{1}, \\ 
&E^{\ul 8} = \sin\chi_{1}\cos\chi_{2}(d\vp_{2}+\alpha_{2}dt), 
\ \ E^{\ul 9} = \sin\chi_{1}\sin\chi_{2}(d\vp_{3}-\alpha_{2}dt).
\eea
On classical solution
\bea
& E^{\ul 0}_{0} = \cosh\sigma, \ \ E^{\ul 1}_{1} = 1, \ \ E^{\ul 2}_{2}=\sinh\sigma, \ \ E^{\ul 3}_{3} = 0, 
\ \ E^{\ul 4}_{4} =  0 \\
& E^{\ul 5}_{5} = 1, \ \ E^{\ul 6}_{6} = 0, \ \ E^{\ul 7}_{7} = 1, \ \ E^{\ul 8}_{8} = 0, 
\ \ E^{\ul 9}_{9} = 0,
\eea
and the non zero induced 2-bein $e^{\ul m}_{a}$ are 
\be
e^{\ul 0}_{0} = \cosh\sigma, \ \ \ e^{\ul 1}_{1} = 1 \qquad \to \qquad 
\rho_{0} = \cosh\sigma\, \Gamma_{\ul 0}, \ \ \ \rho_{1} = \Gamma_{\ul 1}.
\ee

\paragraph{Spin connection and twists}

The spin connection $\omega^{\ul{mn}}_{\mu}$ is obtained from the Cartan equation 
\be
dE^{\ul m}+\omega^{\ul{mn}}\wedge E^{\ul n}=0, \qquad E^{\ul n} = E^{\ul n}_{\mu}dX^{\mu}.
\ee
The twist in $\psi_{3}$ has a peculiar effect  \cite{Tseytlin:1995zv}. Let us follow the $\alpha_{1}$ part in 
\ba
dE^{\ul 4} &= -\alpha_{1}\sinh\rho\sin\psi_{1}\cos\psi_{2}\, d\psi_{2}\wedge dt+\cdots = -\alpha_{1}\cos\psi_{2}\, E^{\ul 3}\wedge dt+\cdots \lp
= -\omega^{\ul{43}}_{0}\wedge E^{\ul 3}+\cdots, \qquad \omega^{\ul{43}}_{0} = -\alpha_{1}\cos\psi_{2}+\cdots,
\ea
where dots are terms vanishing on the classical solution. A similar effect is associated with $\alpha_{2}$ twist.
From
\bea
dE^{\ul 8} &=\alpha_{2} \cos\chi_{1}\cos\chi_{2}d\chi_{1}\wedge dt+\cdots = \alpha_{2}\cos\chi_{1}\cos\chi_{2}\,
E^{\ul 5}\wedge dt+\cdots, \\
dE^{\ul 9} &=- \alpha_{2} \sin\chi_{1}\cos\chi_{2}d\chi_{2}\wedge dt+\cdots = -\alpha_{2}\cos\chi_{2}
E^{\ul 6}\wedge dt+\cdots,
\eea
we get 
\be
\omega^{\ul{85}}_{0} = \alpha_{2}\cos\chi_{1}\cos\chi_{2}+\cdots, \qquad
\omega^{\ul{96}}_{0} = -\alpha_{2}\cos\chi_{2}+\cdots,
\ee
both non vanishing on classical solution.
The other relevant spin connection components can be computed directly for the induced metric
\be
de^{\ul m}+\omega^{\ul{mn}}\wedge e^{\ul n}=0, \qquad e^{\ul n} = e^{\ul n}_{a}d\xi^{a}.
\ee
Since $e^{\ul 0} = \cosh\sigma d\tau$ and $e^{\ul 1}=d\sigma$, the unique case is 
\be
d(\cosh\sigma d\tau)+\omega^{\ul{01}}\wedge d\sigma = 0 \qquad \to \qquad \omega^{\ul{01}} = \sinh\sigma\, d\tau.
\ee
The kinetic part of the action is thus
\ba
L_{F}^{\rm kin} &= 2i\,\sqrt{g}g^{ab}\bar\theta \rho_{a}(\partial_{b}+\tfrac{1}{4}\omega^{\ul{mn}}_{b}\Gamma_{\ul{mn}})\theta 
= 2i\,\bar\theta\bigg[-\frac{1}{\cosh\sigma}\rho_{0}(\partial_{0}+\tfrac{1}{4}\omega^{\ul{mn}}_{0}\Gamma_{\ul{mn}})
+\cosh\sigma\rho_{1}\partial_{1}\bigg]\theta \lp
= 2i\bar\theta[- \Gamma_{\ul 0}(\partial_{0}-\tfrac{1}{2}\alpha_{1}\Gamma_{\ul{43}}
+\tfrac{1}{2}\alpha_{2}\Gamma_{\ul{85}}-\tfrac{1}{2}\alpha_{2}\Gamma_{\ul{96}})
+\cosh\sigma\Gamma_{\ul 1}\partial_{1}+\tfrac{1}{2}\sinh\sigma  \Gamma_{\ul 1}]\, \theta.
\ea
Let us finish by evaluating the mass terms.
In the conventions of \cite{Metsaev:1998it}, the 5-form is 
\be
F_{5} = 2\,(1+\star)\vol_{AdS_{5}} = 2(E^{\ul 0}\wedge \cdots\wedge E^{\ul 4}+E^{\ul 5}\wedge \cdots \wedge E^{\ul 9}),
\ee
where twists are inside three of the vielbeins. Slashing it with curved $\Gamma$ matrices gives
\ba
F_{\mu_{1}\dots\mu_{5}}\Gamma^{\mu_{1}\dots\mu_{5}} &= 2\cdot 5!\,(E^{\ul 0}_{0}\cdots E^{\ul 4}_{4}\, \Gamma^{01234}
+(01234\to 56789)) = 2\cdot 5!\,(\Gamma^{\ul{01234}}+\Gamma^{\ul{56789}}) \lp
= 2\cdot 5!\,\Gamma^{\ul{01234}}(1-\wh\Gamma), \qquad \wh\Gamma = \Gamma^{\ul{012\cdots 9}}.
\ea
Chirality constraint is $\wh\Gamma\theta=\theta$, so 
\be
\mc F_{a}\rho_{b}\theta = -\frac{1}{4}\Gamma^{\ul{01234}}\rho_{b}(1+\wh\Gamma)\theta = 
-\frac{1}{2}\Gamma^{\ul{01234}}\rho_{b}\theta,
\ee
and
\be
L_{F}^{\rm flux} = i\eps^{ab}\bar\theta \rho_{a}\Gamma^{\ul{01234}}\rho_{b}\theta.
\ee
This is 
\ba
L_{F}^{\rm flux} &= i\eps^{ab}\bar\theta \rho_{a}\Gamma^{\ul{01234}}\rho_{b}\theta =i\cosh\sigma
\bar\theta(\Gamma_{\ul 0}\Gamma^{\ul{01234}}\Gamma_{\ul 1}-\Gamma_{\ul 1}\Gamma^{\ul{01234}}\Gamma_{\ul 0})\theta
= -2i\cosh\sigma
\bar\theta\, \Gamma^{\ul{234}}\,\theta,
\ea
and
\be
\la{4.50}
L_{F} = 2i\bar\theta[\Gamma^{\ul 0}(\partial_{0}-\tfrac{1}{2}\alpha_{1}\Gamma_{\ul{43}}
+\tfrac{1}{2}\alpha_{2}\Gamma_{\ul{85}}-\tfrac{1}{2}\alpha_{2}\Gamma_{\ul{96}})+\cosh\sigma\Gamma^{\ul 1}\partial_{1}+\tfrac{1}{2}\sinh\sigma  \Gamma^{\ul 1}
-\cosh\sigma\,\Gamma^{\ul{234}}]\, \theta.
\ee
The last term is a ``$\sigma_{3}$'' constant mass term since $\cosh\sigma=\sqrt g$. In other words, this is same as in \cite{Drukker:2000ep}
up to the coupling to the constant gauge field (all $\pm$ signs are uncorrelated)
\be
L_{F} =2\sqrt{g} \bar\theta[i\,\rho^{a}\wh\nabla_{a}+ \tfrac{1}{2}(\pm\alpha_{1}\pm\alpha_{2}\pm \alpha_{2})
\slashed{A}\pm 1]\theta,\qquad A = (0,1).
\ee
Signs are eigenvalues of commuting $i\Gamma^{\ul{34}}, i\Gamma^{\ul{234}}, i\Gamma^{\ul{85}}, i\Gamma^{\ul{96}}$. 
Since $\alpha_{1}=1$ we have $8+8$ possibilities, with mass $M=\pm 1$ and charge
\be
\pm\tfrac{1}{2}, \pm\tfrac{1}{2}, \pm(\tfrac{1}{2}+\alpha_{2}), \pm(\tfrac{1}{2}-\alpha_{2}),
\ee
corresponding to the 8 fermionic terms in (\ref{4.60}).
Notice also 
that the mass value is the right one
\be
|M| = 1 =  \tfrac{3}{2}-\tfrac{1}{2} = h-\tfrac{1}{2}.
\ee
%

\subsection{Reproducing the line index}

Fields coupled to a constant gauge field in $\tau$ direction receive a shift in the mode number of their Fourier expansion on the 
Euclidean time circle. Following notation of \cite{Beccaria:2023sph} we denote by $\kappa$ this shift. The 
spectrum and shifts of all fluctuation fields are shown in Table \ref{tab:1}.
\begin{table}[H]
\be
\def\arraystretch{1.3}
\begin{array}{ccc|ccc|ccc}
\toprule
\textsc{Fluctuation} & Y_{1} & Z,\bar Z & A_{2} & W_{+}, \overline W_{+} & W_{-}, \overline W_{-} & \psi_{1,2} & \psi_{3,4} & \psi_{5,6,7,8} \\
\midrule
h  & 2 & 2 & 1 & 1 & 1 & \frac{3}{2}& \frac{3}{2}& \frac{3}{2}\\
\kappa & 0 & \alpha_{1} & 0 & \alpha_{2} & -\alpha_{2} & \frac{1}{2}+\alpha_{2} & \frac{1}{2}-\alpha_{2} & 0\\
\bottomrule
\end{array}\notag
\ee
\caption{Conformal dimension and $\kappa$-shifts for the scalar fluctuations in $AdS_{5}$, $S^{5}$, and for the eight fermionic modes.}
\la{tab:1}
\end{table}
Adapting to $AdS_{2}$ the discussion in  \cite{Beccaria:2023sph}, 
the partition function of a scalar with conformal dimension $h$ in thermal $AdS_{2}$
gives the following one-loop determinant, \cf also Appendix \ref{app:der},
\be
\la{4.54}
\Gamma^{(h, \kappa)} = \frac{1}{2}\log\det \Delta^{(h,\kappa)} = 
\beta E_{c}^{(h,\kappa)}-\frac{1}{2}\sum_{n=1}^{\infty}\frac{1}{n}\frac{q^{n h}(q^{n\kappa}+q^{-n\kappa})}{1-q^{n}}, 
\qquad q=e^{-\beta},
\ee
where $E_{c}$ is the supersymmetric Casimir energy contribution. Notice that this can be written 
\be
\la{4.55}
\exp(-\Gamma^{(h,\kappa)}) = e^{-\beta E_{c}^{(h,\kappa)}}\PE\bigg[-\frac{1}{2}\frac{q^{h}(q^{\kappa}+q^{-\kappa})}{1-q}\bigg].
\ee
The expression of the supersymmetric Casimir energy $E_{c}$ is, see  \cite{Giombi:2014yra} and Appendix \ref{app:Casimir},
\bea
\la{4.56}
E_{c} &= \frac{1}{2}\zeta_{E}(-1), \qquad \zeta_{E}(z) = \frac{1}{\Gamma(z)}\int_{0}^{\infty}d\beta \beta^{z-1}Z(\beta),\\
Z &= \frac{1}{2}\wt Z(h+\kappa)+\frac{1}{2}\wt Z(h-\kappa), \qquad \wt Z(\Delta) = \frac{q^{\Delta}}{1-q}. 
\eea
We evaluate the integral in terms of Hurwitz zeta function \footnote{This is value of the integral from definition of Hurwitz zeta function or as Mellin 
transform.  Alternatively, 
one may expand $1/(1-e^{-\beta})$ and gets the same from $ \sum_{n=0}^{\infty}(n+\Delta)^{-z} = \zeta(z,\Delta)$.}
\be
\frac{1}{\Gamma(z)}\int_{0}^{\infty}d\beta \beta^{z-1}\frac{e^{-\beta \Delta}}{1-e^{-\beta}} =  \zeta(z,\Delta),
\ee
and then 
\be
 \frac{1}{2}\zeta'(-1,\Delta) = \frac{1}{24}(-1+6\Delta-6\Delta^{2}).
\ee
Taking half the sum with $\Delta = h\pm\kappa$ gives 
\be
\la{4.59}
E_{c}^{(h, \kappa)} = \frac{1}{24}(-1+6h-6h^{2}-6\kappa^{2}).
\ee
From the data in Table \ref{tab:1}, we need to compute the combination \footnote{Fermions have here periodic boundary condition in 
Euclidean time. Their contribution to $\Gamma_{\oneloop}$ is same as for scalars, up to a minus sign.}
\ba
\la{4.60}
\Gamma_{\rm one-loop} &=  \Gamma^{(2,0)}+2\Gamma^{(2,\alpha_{1})}+\Gamma^{(1,0)}+4\Gamma^{(1,\alpha_{2})}
- [2\Gamma^{(\frac{3}{2},\frac{1}{2}+\alpha_{2})}+2 \Gamma^{(\frac{3}{2},\frac{1}{2}-\alpha_{2})}
+4 \Gamma^{(\frac{3}{2},\frac{1}{2})}], 
\ea
with a similar sum over fields for total $E_{c, \rm one-loop}$. 
Comparing with (\ref{3.3}), we  replace in (\ref{4.60})
\be
\la{4.61}
\alpha_{1}=1, \qquad q^{\alpha_{2}} = \eta.
\ee
This gives, \cf (\ref{4.55}),
\be
\I_{\F}( \eta; q) = \exp(-\Gamma_{\rm one-loop}) = \PE[-q^{2}+(\eta+\eta^{-1})\,q] = \frac{1-q^{2}}{(1-\eta q)(1-\eta^{-1}q)},
\ee
in agreement with the single particle index (\ref{2.6}). The total one-loop Casimir contribution vanishes $E_{c, \rm one-loop}=0$.
We have thus reproduced the large $N$ line index by a semiclassical computation in twisted background.

\section{Giant graviton correction to $\I_{\F}(\eta; q)$}
\la{sec:line-giant}

According to the proposal in \cite{Imamura:2024lkw}, the  leading large $N$ corrections to the line index are captured by fluctuations
of two half-infinite  fundamental (F) strings with worldsheet ending on one of the two Wilson lines and D3 giant graviton.
The exact expression in (\ref{2.8}) shows that D3 brane and  F string fluctuations decouple at this order. The role of the 
D3 giant is to assign specific boundary conditions for F string fluctuations.

The D3 brane worldvolume is $S^{1}_{\beta}\times S^{3}$ where 
$S^{3}\subset S^{5}$ and is at $\rho=0$. 
The F string is pointlike in $S^{5}$ and three fluctuations are along $S^{3}$ of the D3 brane. These should obey 
Neumann boundary conditions at $\rho=0$. All other fluctuations should have Dirichlet boundary conditions at $\rho=0$.
This way we have two worldsheets each corresponding to an edge on the D3 brane and the other  
on the Wilson line at AdS boundary $\rho\to\infty$, \cf Fig. \ref{fig:1}.

\begin{figure}[H]
\begin{center}
\begin{tikzpicture}[line width=1 pt, scale=0.7, rotate=0,baseline=0]
\draw (0,0) circle (2);
\draw[thin] (-2,0)--(2,0);
\node[right] at (45:2.2) {$\partial AdS_{5} = \mathbb{R}\times S^{3}$};
\node[right] at (2.2,0) {$L$};
\node[left] at (-2.2,0) {$\bar L$};
\node at (0,-2.75) {(a)};
\end{tikzpicture}
\begin{tikzpicture}[line width=1 pt, scale=0.7, rotate=0,baseline=0]
\draw (0,0) circle (2);
\draw[thin] (-2,0)--(2,0);
\draw[fill=white] (0,0) circle (0.1);
\node[right] at (45:2.2) {$\partial AdS_{5} =  \mathbb{R}\times S^{3}$};
\node[right] at (2.2,0) {$L$};
\node[left] at (-2.2,0) {$\bar L$};
\node[above] at (0,0) {D3};
\node at (0,-1.2) {$\partial_{\rho}\Phi=0$};
\draw[thin,-latex] (0,-0.8)--(0.3,0); 
\draw[thin,-latex] (0,-0.8)--(-0.3,0); 
\node at (0,-2.75) {(b)};
\end{tikzpicture}
\end{center}
\caption{
\la{fig:1}
(a) without D3 brane: One $AdS_{2}$ worldsheet with edges ending on Wilson lines;  
(b) in the presence of the D3 brane: Two $AdS_{2}$ worldsheets each having an edge on the 
D3 brane and the other on a Wilson line. In the figure, $\Phi$ is any of the three $S^{5}$ scalar fluctuations along $S^{3}$ of the D3 brane.}
\end{figure}
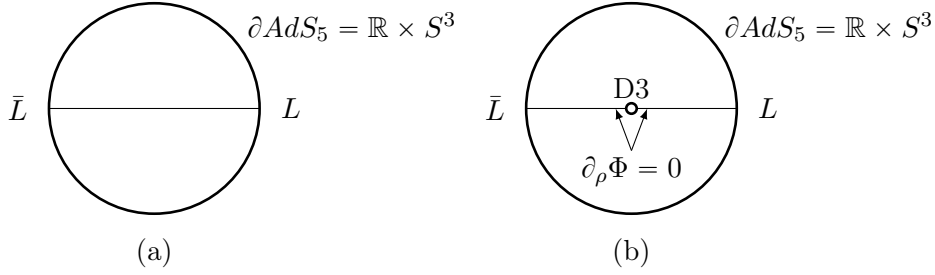

The square of the corresponding one-loop determinant should reproduce the correction in (\ref{2.8}), \ie  
the factor 
\be
\la{5.1}
\exp(-2\Gamma_{\rm one-loop}^{\rm D3}) = \frac{1}{\eta q}\PE[2\eta^{-1}q-2q^{2}].
\ee
This is same as 
\be
\la{5.2}
\Gamma_{\rm one-loop}^{\rm D3} =-\frac{1}{2}\log\big[ \frac{1}{\eta q}\PE[2\eta^{-1}q-2q^{2}]\big]  = -\frac{1}{2}(1+\alpha_{2})\,\beta-\sum_{n=1}^{\infty}\frac{1}{n}
(\eta^{-n}q^{n}-q^{2n}),
\ee
where we used (\ref{4.61}). This suggests that the elusive (and unclear) factor $\frac{1}{\eta q}$ found in
\cite{Imamura:2024lkw} is related to a non-vanishing 
Casimir energy contribution due to reduced supersymmetry in the presence of the D3 brane.

What we now need is to replace the standard effective action $\Gamma$ on thermal $AdS_{2}$ in (\ref{4.54}) by an expression that takes into account only fluctuations modes with the assigned boundary condition on the D3 brane. This is done in next section by 
a normal mode decomposition.

\paragraph{Remark} These are not boundary conditions at  $\rho\to \infty$. All fields will be outside the Breitenlohner-Freedman
window \cite{Breitenlohner:1982jf} and admit a regular quantization 
with $\Phi\sim \exp(-h_{\Phi}\rho)$ at infinity. Instead,
boundary conditions on the D3 brane at $\rho=0$ refer to parity of states under $\rho\to -\rho$. In this respect, at least for scalars, 
even parity states obey Neumann  boundary conditions at the D3 brane, while odd parity states obey Dirichlet boundary conditions.
The case of fermionic fluctuations will be discussed later in full details.

\subsection{One-loop thermal determinant in $AdS_{2}$ and  boundary conditions at $\rho=0$}

\subsubsection{Scalar field}
\la{sec:scalar}

In the normal mode method \cite{Denef:2009kn,Denef:2009yy}, \footnote{
The method was developed for non-trivial black hole background where normal modes are actually dissipative quasi-normal modes.
This is not the case here where we have genuine normal modes corresponding to  Euclidean zero modes with suitable boundary conditions.}
the determinant of kinetic operator for a free scalar field 
in $AdS_{2}$ with dual conformal dimension $h$ (and mass $m^{2}=h(h-1)$ ) is written exploiting Weierstrass representation
\be
\det(-\nabla^{2}+h(h-1)) = e^{-P(h)}\prod_{I}(h-h_{I}),
\ee
where $h_{I}$ are special values of the analytically continued conformal dimension $h$ to the complex plane
that correspond to Euclidean zero modes of the kinetic operator. The entire function $\exp(-P(h))$ can be fixed by  matching the zeta-function regularization at large $h$ to the local heat kernel curvature expansion expression in this limit. 
\footnote{It is
a polynomial in $h$, obtained  by computing a finite number of integrals over spacetime of local curvature invariants.}

In thermal $AdS_{2}$ (and Euclidean time $\tau$) a zero mode has time dependence 
$\sim e^{-\Omega\tau}$ and periodicity in thermal cycle requires
\be
\Omega = i\,\omega_{n}, \qquad \omega_{n} = \frac{2\pi n}{\beta}.
\ee
This means a pole in $\Gamma(1+\frac{i\beta}{2\pi}\Omega)$ in the variable $\Omega$. If frequencies are symmetric under 
$\Omega\to -\Omega$ one has 
\ba
\det(-\nabla^{2}+h(h-1))^{-1} &= e^{P(h)}\prod_{I}\frac{\beta}{4\pi^{2}}\Omega_{I}
\Gamma\bigg(\frac{i\beta}{2\pi}\Omega_{I}\bigg)\Gamma\bigg(-\frac{i\beta}{2\pi}\Omega_{I}\bigg) \lp
= \mc N\, e^{P(h)}\prod_{n>0}\bigg(n+\frac{i\beta}{2\pi}\Omega_{I}\bigg)^{-1}\bigg(n-\frac{i\beta}{2\pi}\Omega_{I}\bigg)^{-1},
\ea
(where we included in $\mc N$ the contribution from constant solutions with zero $\Omega$ and additional UV divergent factors).
This may be written concisely as 
\be
\la{5.6}
\det(-\nabla^{2}+m^{2}) = \mc N\, e^{P(h)}\prod_{n>0, I}(i\omega_{n}(\beta)-\Omega_{I}(h))(-i\omega_{n}(\beta)-\Omega_{I}(h)).
\ee
Normal mode frequencies are found by solving the Klein-Gordon equation and imposing boundary conditions at infinity, corresponding
to $h$. In  the rather special case of $AdS_{2}$, normal mode frequencies  are parametrized by a single non-negative integer \cite{Martin:2019flv,Keeler:2014hba,Keeler:2016wko}
\be
\la{5.7}
\{\Omega_{I}\} = \{\Omega_{p} = \pm(h+p), \ \  p=0,1,2,\dots\}.
\ee
To see this in our locally $AdS_{2}$ specific metric, let us write the Euclidean wave equation as
\be
\cosh\rho\,\partial_{\rho}(\cosh\rho\,\partial_{\rho}\vp)+\partial_{\tau}^{2}\vp-m^{2}\cosh^{2}\rho\,\vp=0, \qquad m^{2}=h(h-1).
\ee
We consider 
\be
\phi(\rho, \tau) = \phi(\rho)\, e^{\Omega\,\tau},
\ee
and thus $\partial_{\tau}^{2}=\Omega^{2}$. The general solution for $\vp(\rho)$ is 
\be
\la{5.10}
\vp(\rho) = (1-\tanh^{2}\rho)^{1/4}\bigg[C_{1}\, P_{\Omega-\frac{1}{2}}^{h-\frac{1}{2}}(\tanh \rho)
+C_{2}\,Q_{\Omega-\frac{1}{2}}^{h-\frac{1}{2}}(\tanh \rho)
\bigg],
\ee
where $P,Q$ are associated Legendre functions. For integer $h\ge 1$,  the  solution vanishing at $\rho\to \infty$ has $C_{1}=0$
and it goes correctly as $e^{-h\rho}$.
At  $r\to 0$ we have then 
\be
\vp(\rho) = C'\bigg[\cos\frac{\pi(h+\Omega)}{2}+\frac{2\Gamma(\frac{2-h+\Omega}{2})\Gamma(\frac{1+h+\Omega}{2})}
{\Gamma(\frac{1-h+\Omega}{2})\Gamma(\frac{h+\Omega}{2})}\, \sin\frac{\pi(h+\Omega)}{2}\, \rho +\mc O(\rho^{2}).
\bigg].
\ee
Notice that this is smooth at $\rho=0$ for any $\Omega$, contrary to what happens in higher dimension,
as pointed out in \cite{Sakai:1984vm}. We now require $\vp(0)=0$ or $\vp'(0)=0$. This condition ensures 
that the resulting spectrum is discrete in spatial direction and not continuous as one would expect in 
a non-compact hyperbolic space, see for instance \cite{Higuchi:2021fxg}. 
More physically, it corresponds to Dirichlet boundary conditions at the two boundaries of $AdS_{2}$. \footnote{
Euclidean $AdS_{2}$ (with unit radius) is the one sheet $X_{0}\ge 1$ of   $X_{0}^{2}-X_{E}^{2}-X_{1}^{2}=1$
covered once by coordinates $(X_{0},X_{1},X_{E}) = (\sec\sigma\cosh\tau, \tan\sigma, \sec\sigma\sinh\tau)$
with $\sigma\in(-\pi/2,\pi/2)$ and $\tau\in\mathbb R$. The induced metric is 
$ds^{2} = -dX_{0}^{2}+dX_{E}^{2}+dX_{1}^{2} = \frac{1}{\cos^{2}\sigma}(d\tau^{2}+d\sigma^{2})$,
with two boundaries at $\sigma = \pm \frac{\pi}{2}$. With the change of variable $\sinh\rho=\tan\sigma$ 
the metric becomes $ds^{2} = \cosh^{2}\rho\, d\tau^{2}+d\rho^{2}$
and we should take here $\rho\in(-\infty, \infty)$, see for instance \cite{Aharony:1999ti}. This is in contrast with higher dimensional AdS
where $X_{1}\to X_{i} = R \sinh\rho \ \Omega_{i}$ with $\sum_{i}\Omega_{i}^{2}=1$ and sign of $X_{i}$ can be any while
$\rho\in \mathbb R^{+}$ is radius. 
}
This  gives (\ref{5.7})
where $\vp$ is even/odd for even/odd values of $p$. 

Using these  frequencies in (\ref{5.6}) gives quickly \footnote{We absorbing in $P(h)$ various divergent factors. 
These may be treated carefully, but the aim of this derivation is to show how the non-Casimir part of (\ref{4.54}) emerges.
The remaining Casimir contribution can be quickly fixed later as explained in Appendix \ref{app:Casimir}.
}
\ba
\la{5.12}
\prod_{n>0, p\ge 0} & (i\omega_{n}(\beta)-\Omega_{p}(h))(-i\omega_{n}(\beta)-\Omega_{p}(h))= \prod_{n>0, p\ge 0}[\tfrac{2\pi i n}{\beta}-(p+h)]^{2}
[-\tfrac{2\pi i n}{\beta}-(p+h)]^{2}\lp
 \to  \prod_{n>0, p\ge 0} \bigg(1+\frac{\beta^{2}(p+h)^{2}}{4\pi^{2}n^{2}}\bigg)^{2}.
\ea
We now use
\be
\prod_{n>0}\bigg(1+\frac{a^{2}}{n^{2}}\bigg) = \frac{e^{\pi a}}{2\pi a}(1-e^{-2\pi a}),
\ee
and get 
\be
\det(-\nabla^{2}+m^{2}) = \mc N\, \prod_{p\ge 0}(1-q^{p+h})^{2}, \qquad q = e^{-\beta}.
\ee
Hence, expanding in $q$,  summing over $p$, and denoting by a prime the non-Casimir  part
\ba
\la{5.15}
\frac{1}{2}\log\det(-\nabla^{2}+m^{2})' &= \sum_{p\ge 0}\log(1-q^{p+h}) = -\sum_{p\ge 0}
\sum_{n>0}\frac{1}{n}q^{n(p+h)} 
= -\sum_{n>0}\frac{1}{n}\frac{q^{nh}}{1-q^{n}},
\ea
which reproduces   (\ref{4.54}) in neutral case $\kappa=0$.

\paragraph{Charged case}

For a charged scalar field coupled to a constant gauge field
in $\tau$ direction the discussion is almost unchanges. If $\partial_{\tau}\to \partial_{\tau}-\kappa$ in the wave equation, this means that 
$\Omega$ enters the previous calculation with the replacement $\Omega\to \Omega-\kappa$. Normal mode frequencies are then 
\be
\Omega_{p} = \kappa\pm (h+p), \quad p=0,1,2,\dots.
\ee
Thus,
\ba
\prod_{n>0, I} & (i\omega_{n}(\beta)-\Omega_{I}(h))(-i\omega_{n}(\beta)-\Omega_{I}(h)) \lp
= \prod_{p}
[\tfrac{2\pi i n}{\beta}-(h+\kappa+p)][-\tfrac{2\pi i n}{\beta}-(h+\kappa+p)]
[\tfrac{2\pi i n}{\beta}+h-\kappa+p][-\tfrac{2\pi i n}{\beta}+h-\kappa+p] ,
\ea
and this gives the non-Casimir part of  (\ref{4.54}) for a full complex field (two real components)
\ba
\la{5.18}
\log\det(-\nabla^{2}+m^{2})' &= -2\times \frac{1}{2}\sum_{n>0}\frac{1}{n}\frac{q^{n(h+\kappa)}+q^{n(h-\kappa)}}{1-q^{n}}.
\ea

\paragraph{Neumann/Dirichlet boundary conditions}

Now, suppose we keep in (\ref{5.7}) only modes with even/odd $p$, corresponding to Neumann/Dirichlet conditions at $\rho=0$. 
This is a restriction on $p$ summation in \eg (\ref{5.15}) and for even/odd $p$ we get 
\bea
\text{even}\ p: &  \sum_{k\ge 0}\sum_{n>0}\frac{1}{n}q^{n(2k+h)} 
= -\sum_{n>0}\frac{1}{n}\frac{q^{nh}}{1-q^{2n}},\\
\text{odd}\ p: &  \sum_{k\ge 0}\sum_{n>0}\frac{1}{n}q^{n(2k+1+h)} 
= -\sum_{n>0}\frac{1}{n}\frac{q^{n(h+1)}}{1-q^{2n}}.
\eea
In conclusion, we can split (\ref{4.54}) into contributions from fields with Neumann/Dirichlet boundary conditions as 
\bea
\la{5.20}
\Gamma^{(h, \kappa)}_{\rm N} & = 
\beta E_{c, \rm N}^{(h,\kappa)}-\frac{1}{2}\sum_{n=1}^{\infty}\frac{1}{n}\frac{q^{n h}(q^{n\kappa}+q^{-n\kappa})}{1-q^{2n}},\\
\Gamma^{(h, \kappa)}_{\rm D} & = 
\beta E_{c, \rm D}^{(h,\kappa)}-\frac{1}{2}\sum_{n=1}^{\infty}\frac{1}{n}\frac{q^{n (h+1)}(q^{n\kappa}+q^{-n\kappa})}{1-q^{2n}}.
\eea
The Casimir energies can be computed as before and (\ref{4.59}) splits into
\ba
\la{5.21}
E_{c, \rm N}^{(h, \kappa)} &= \frac{1}{24}(-2+6h-3h^{2}-3\kappa^{2}), \qquad
E_{c, \rm D}^{(h, \kappa)} = \frac{1}{24}(1-3h^{2}-3\kappa^{2}).
\ea
In the following, for charged scalars it will be  convenient to separate the two contributions with $\pm \kappa$
in (\ref{5.18})
and write (hat denoting one sign of $\kappa$)
\ba
\la{5.22}
\wh \Gamma^{(h, \kappa)}_{\rm N} & = 
\beta \wh E_{c, \rm N}^{(h,\kappa)}-\frac{1}{2}\sum_{n=1}^{\infty}\frac{1}{n}\frac{q^{n (h+\kappa)}}{1-q^{2n}},\qquad
\wh\Gamma^{(h, \kappa)}_{\rm D}  = 
\beta \wh E_{c, \rm D}^{(h,\kappa)}-\frac{1}{2}\sum_{n=1}^{\infty}\frac{1}{n}\frac{q^{n (h+1+\kappa)}}{1-q^{2n}},
\ea
and 
\ba
\la{5.23}
\wh E_{c, \rm N}^{(h, \kappa)} &= \frac{1}{48}(-2+6h-3h^{2}+6\kappa(1-h)-3\kappa^{2}), \qquad
\wh E_{c, \rm D}^{(h, \kappa)} = \frac{1}{48}(1-3h^{2}-6\kappa\, h-3\kappa^{2}).
\ea

\subsubsection{Spin-$\frac{1}{2}$ field}
\la{sec:fermion}

Let us now consider a spin-$\frac{1}{2}$ fermion in our $AdS_{2}$ induced metric (\ref{4.14}). 
Zero modes are solutions of $K\theta=0$ where ($\tau_{i}$ are Pauli matrices)
\be
K = \tau_{1}\partial_{\tau}+\tau_{3}(\cosh\rho \partial_{1}+\tfrac{1}{2}\sinh\rho)-i M\tau_{2}\cosh\rho.
\ee
We may redefine
\be
\theta \equiv \binom{\theta_{1}}{\theta_{2}}= (\cosh\rho)^{-1/2}\binom{\chi_{1}}{\chi_{2}},
\ee
to get 
\be
K = \frac{1}{\cosh\rho}\tau_{1}\partial_{\tau}+\tau_{3}\partial_{1}-i M\tau_{2}.
\ee
Let us find the  normal mode frequencies by considering
\be
\theta(\rho, \tau) = \theta(\rho)\, e^{\Omega\tau},
\ee
and solving
\be
\bigg(\frac{\Omega}{\cosh\rho}\tau_{1}+\tau_{3}\partial_{1}-i M\tau_{2}\bigg)\binom{\chi_{1}}{\chi_{2}}=0.
\ee
\ba
\la{5.29}
& \frac{\Omega}{\cosh\rho}\chi_{2}+\chi_{1}'-M\chi_{2}=0, \qquad
 \frac{\Omega}{\cosh\rho}\chi_{1}-\chi_{2}'+M\chi_{1}=0.
\ea
Let us set $M=h-\frac{1}{2}$ and 
\be
\binom{\chi_{1}(\rho)}{\chi_{2}(\rho)} = U\,\binom{A(\rho)}{B(\rho)}, \qquad U=\begin{pmatrix} i & 1 \\ -i & 1\end{pmatrix}.
\ee
The two equations (\ref{5.29}) take the slightly more decoupled form 
\be
A'+(h-\frac{1}{2})A-i\frac{\Omega}{\cosh\rho}B=0, \qquad 
-B'+(h-\frac{1}{2})B+i\frac{\Omega}{\cosh\rho}A=0.
\ee
In particular, we have 
\be
\la{5.32}
B = -i\frac{\cosh\rho}{\Omega}\bigg(A'+(h-\tfrac{1}{2})A\bigg),
\ee
and the second order differential equation for $A(\rho)$
\be
A''+\tanh\rho A'+\bigg[\frac{\Omega^{2}}{\cosh^{2}\rho}-(h-\tfrac{1}{2})(h-\tfrac{1}{2}-\tanh\rho)\bigg]A=0.
\ee
The solution corresponding to $\theta(\rho)$ vanishing at $\rho\to \infty$ as $e^{-\rho h}$ is -- using (\ref{5.32}) and up to an overall
normalization constant --
\bea
A(\rho) &= \frac{(1-\tanh\rho)^{h/2}(1+\tanh\rho)^{\frac{1-h}{2}}}{(1-\tanh^{2}\rho)^{1/4}}\, 
{}_{2}F_{1}(-\Omega, \Omega, h, \tfrac{1-\tanh\rho}{2}), \\
B(\rho) &= -\frac{i\Omega }{2h} \frac{(1-\tanh\rho)^{-1+h/2}(1+\tanh\rho)^{-\frac{1}{2}-\frac{h}{2}}}{(\cosh\rho)^{5/2}}\, 
{}_{2}F_{1}(1-\Omega, 1+\Omega, 1+h, \tfrac{1-\tanh\rho}{2}).
\eea
The explicit form of $\theta(\rho)$ is then 
\be
\theta(\rho) = \frac{i}{\sqrt{\cosh\rho}}\, U\, \binom{A(\rho)}{B(\rho)},
\ee
and one can check that it is real. Besides, at small $\rho$ we find for $h=\frac{3}{2}$ the expansions
\bea
\theta_{1}(\rho) &= -\frac{1}{1+2\Omega}\bigg(\cos\frac{\pi\Omega}{2}+\sin\frac{\pi\Omega}{2}\bigg)
+\frac{\Omega-1}{2\Omega-1}\bigg(\cos\frac{\pi\Omega}{2}-\sin\frac{\pi\Omega}{2}\bigg)\, \rho+\mc O(\rho^{2}), \\
\theta_{2}(\rho) &= \frac{1}{1-2\Omega}\bigg(\cos\frac{\pi\Omega}{2}-\sin\frac{\pi\Omega}{2}\bigg)
-\frac{\Omega+1}{2\Omega+1}\bigg(\cos\frac{\pi\Omega}{2}+\sin\frac{\pi\Omega}{2}\bigg)\, \rho+\mc O(\rho^{2}).
\eea
We get even/odd components according to 
\bea
\la{5.37}
&\theta_{1}(0)\neq 0 \ \text{and}\ \theta_{2}(0)=0\ \ :\ \Omega \,\text{mod}\, 4 = \tfrac{1}{2}, -\tfrac{3}{2}, \quad \Omega\neq \tfrac{1}{2}, \\
&\theta_{1}(0)= 0 \ \text{and}\ \theta_{2}(0)\neq 0\ \ :\ \Omega \,\text{mod}\, 4 = -\tfrac{1}{2}, \tfrac{3}{2}, \quad \Omega\neq -\tfrac{1}{2}.
\eea
%
These values correspond to the same formula we found in 
bosonic case, \ie  $\Omega =\pm( h+p)$, $p=0,1,2,\dots$ where here $h=\tfrac{3}{2}$. 
The first line in (\ref{5.37}) corresponds to odd modes $p\in 2\mathbb N+1$, while the second line is for even modes $p\in 2\mathbb N$.
If we refer to the parity of $\theta_{1}$, we can say that even modes are of Dirichlet type, while odd modes are of Neumann type.
The second component $\theta_{2}$ has opposite behaviour.

\subsection{Boundary conditions on D3 giant and supersymmetry}

Let us now use supersymmetry to consistently pair the parities of bosonic and fermionic modes.
Using notation of \cite{Imamura:2024lkw} for worldsheet fluctuations, the states in short multiplet of $OSp(4^{*}|4)$ in (\ref{3.7}) are denoted
\be
\la{5.38}
\vp(1,\bm{1}, \bm{5})+\psi(\tfrac{3}{2}, \bm{2}, \bm{4})+\phi(2, \bm{3}, \bm{1}).
\ee
In our notation, bosonic fluctuation are 
\be
S^{5}: \vp \equiv (A_{2}, W_{+}, \overline W_{+}, W_{-}, \overline W_{-}), \qquad AdS_{5}: \phi \equiv (Y_{1}, Z, \bar Z).
\ee
In the presence of the D3 brane, the R-symmetry reduces to 
 $SO(5)_{R}\to SO(2)_{R}\times SO(3)_{R}$ and the five $S^{5}$ fluctuations split into 
$3+1+1$ scalars 
$\vp = (\vp_{0}, \vp_{+}, \vp_{-})$ transforming respectively  in the $\bm{3}, \bm{1}, \bm{1}$ of $SO(3)_{R}$.
Under the supersymmetry preserved by the D3 brane the superconformal representation (\ref{5.38}) splits into small representations
summarized in Table \ref{tab:2} \footnote{This same as Table 1 in \cite{Imamura:2024lkw} adapted to our notation.}
\begin{table}[H]
\be
\def\arraystretch{1.3}
\begin{array}{ccccccc}
\toprule
\textsc{Fluctuation} & h & SU(2) & \sfR & SO(3)_{R}  && \text{mode} \\
\midrule
\vp_{-} & n-1 & \bm{1} & -1 & \bm{1} & n\ge 2 & n-2\\
\psi_{-} & n-\frac{1}{2} & \bm{2} & -\frac{1}{2} & \bm{2}& n\ge 2  & n-2\\
\vp_{0} & n & \bm{1} & 0 & \bm{3}  & n\ge 2 & n-1\\
\phi & n & \bm{3} & 0 & \bm{1} & n\ge 1 & n-2\\
\psi_{+} & n+\frac{1}{2} & \bm{2} & +\frac{1}{2} & \bm{2} & n \ge 1 & n-1\\
\vp_{+} & n+1 & \bm{1} & +1 & \bm{1} & n \ge 0 & n\\
\bottomrule
\end{array}\notag
\ee
\caption{Fluctuation modes in small superconformal representation in the presence of the D3 brane. }
\la{tab:2}
\end{table}
The structure of the small superconformal representation is discussed in full details in Appendix \ref{app:osp} and reads
\be
\la{5.40}
\begin{tikzpicture}[line width=1 pt, scale=0.7, rotate=0,baseline=0]
\node at (0,0) {$\vp_{-}$};
\node at (2.5,0) {$\psi_{-}$};
\node at (6,0.7) {$B^{+}\vp_{0}$};
\node at (6,-0.7) {$\phi$};
\node at (10,0) {$B^{+}\psi_{+}$};
\node at (14,0) {$(B^{+})^{2}\vp_{+}$};
\draw[->,thin] (0+0.7,0)--(2.5-0.7,0);
\node at (1.25,0.3) {$\text{\scriptsize Q}$};
\draw[->,thin] (11,0)--(12.5,0);
\node at (11.75,0.3) {$\text{\scriptsize Q}$};
\draw[->,thin] (3,0)--(5,0.5);
\node[above] at (4,0.25) {$\text{\scriptsize Q}$};
\node[below] at (4,-0.25) {$\text{\scriptsize Q}$};
\draw[->,thin] (3,0)--(5,-0.5);
\draw[->,thin] (7,0.5)--(9,0.2);
\draw[->,thin] (7,-0.5)--(9,-0.2);
\node[above] at (8,0.25) {$\text{\scriptsize Q}$};
\node[below] at (8,-0.25) {$\text{\scriptsize Q}$};
\end{tikzpicture}
\ee
where $Q$ are the supercharges preserved by the D3 brane and $B^{+}$ is the raising operator of  $SL(2, \mathbb R)$
generating conformal descendants. Each application of $B^{+}$ shifts forward the mode number.
The mode of each state is $h-h_{0}$ where $h_{0}$ are the dilatation eigenvalues in (\ref{5.38}), so $h_{0}=1$ for $\vp_{0}, \vp_{\pm}$, $h_{0}=\frac{3}{2}$
for $\psi_{\pm}$, and $h_{0}=2$ for $\phi$.
Bosonic boundary conditions on the D3 brane are 
\be
\vp_{0}\,:\,\text{Neumann}, \qquad \vp_{\pm}, \phi\ :\ \text{Dirichlet},
\ee
corresponding to three $S^{5}$ fluctuations longitudinal to the D3 brane, while all other bosonic fluctuations
(two in $S^{5}$ and three in $AdS_{5}$)  are transverse. As already noted in \cite{Imamura:2024lkw} this requires
$n$ to be odd in the full multiplet in Table \ref{tab:2}.

Fermionic fluctuations $\psi_{-}$ have then odd mode number and are of Neumann type according to the discussion after (\ref{5.37}).
Instead, fluctuations $\psi_{+}$ have even mode number and are thus of Dirichlet type.
Let us now consider the fermion determinant computed restricting  $\theta_{1}$ to have  Neumann b.c.
in the presence of the constant gauge field along $\tau$ direction.
The product in (\ref{5.6}) reads (recall exclusion of $\Omega=\pm 1/2$, denoted with a prime)
\ba
& \prod_{n\in \mathbb Z,p\in\mathbb Z}[\tfrac{2\pi i n}{\beta}-(\tfrac{1}{2}+4p+\kappa)]'
[\tfrac{2\pi i n}{\beta}-(-\tfrac{3}{2}+4p+\kappa)] \lp
\prod_{n\in \mathbb Z,p\ge 0}[\tfrac{2\pi i n}{\beta}-(\tfrac{3}{2}+4p+3+\kappa)][\tfrac{2\pi i n}{\beta}+(\tfrac{3}{2}+4p+2-\kappa)]
[\tfrac{2\pi i n}{\beta}-(\tfrac{3}{2}+4p+1+\kappa)][\tfrac{2\pi i n}{\beta}+(\tfrac{3}{2}+4p-\kappa)] \lp
= \prod_{n\in \mathbb Z,p\ge 0}[\tfrac{2\pi i n}{\beta}-(\tfrac{3}{2}+2p+1+\kappa)][\tfrac{2\pi i n}{\beta}+(\tfrac{3}{2}+2p-\kappa)].
\ea
Thus, we have same expression as the $+\kappa$ contribution for a half-boson (with $h=\frac{3}{2}$) with Dirichlet boundary conditions plus the $-\kappa$ 
contribution for a half-boson with Neumann boundary condition.
The corresponding total contributions is  
\be
\la{5.43}
\wh \Gamma^{(h,\kappa)}_{\rm D}+\wh \Gamma^{(h,-\kappa)}_{\rm N}.
\ee
Similarly, we can consider the fermion determinant restricting $\theta_{1}$ to have Dirichlet boundary condition. In this case we need the product
\ba
 \prod_{n\in \mathbb Z,p\in\mathbb Z} & [\tfrac{2\pi i n}{\beta}-(-\tfrac{1}{2}+4p+\kappa)]'
[\tfrac{2\pi i n}{\beta}-(\tfrac{3}{2}+4p+\kappa)] \lp
= \prod_{n\in \mathbb Z,p\ge 0}[\tfrac{2\pi i n}{\beta}-(\tfrac{3}{2}+2p+\kappa)][\tfrac{2\pi i n}{\beta}+(\tfrac{3}{2}+2p+1-\kappa)],
\ea
that gives
\be
\la{5.45}
\wh \Gamma^{(h,\kappa)}_{\rm N}+\wh \Gamma^{(h,-\kappa)}_{\rm D}.
\ee
To summarize, the structure of the small representation in Table \ref{tab:2} implies that fermions will contribute 
by the sum of the two combinations in (\ref{5.43}) and (\ref{5.45}).

\subsection{Total one-loop contribution in the presence of the D3 brane}

The previous expression (\ref{4.60}) valid in the absence of the D3 brane 
may be written using for fermionic contributions depending on $\alpha_{2}$
the split expressions  (\ref{5.22}) 
\ba
\la{5.46}
\Gamma_{\rm one-loop} &=  \underbrace{\Gamma^{(2,0)}+2\Gamma^{(2,1)}}_{AdS_{5}}
+\underbrace{\Gamma^{(1,0)}+4\Gamma^{(1,\alpha_{2})}}_{S^{5}}\lp
- 2\,[
\wh\Gamma^{(\frac{3}{2},\frac{1}{2}+\alpha_{2})}+\wh\Gamma^{(\frac{3}{2},-\frac{1}{2}-\alpha_{2})}
+ \wh\Gamma^{(\frac{3}{2},\frac{1}{2}-\alpha_{2})}+ \wh\Gamma^{(\frac{3}{2},-\frac{1}{2}+\alpha_{2})}
+2\wh\Gamma^{(\frac{3}{2},\frac{1}{2})}+2\wh\Gamma^{(\frac{3}{2},-\frac{1}{2})}], 
\ea
where we recall that for the purpose of counting states $\wh\Gamma$ gives 1/2 while $\Gamma$ gives 1 so we have $8_{B}+8_{F}$
contributions.

We now compute (\ref{5.46}) by using Dirichlet boundary conditions for the 3 AdS scalar fluctuations 
and for 2 of the $S^{5}$ ones. The remaining 3 fluctuations in $S^{5}$ have Neumann boundary condition.
For fermions, we don't have a natural $AdS_{5}+S^{5}$ splitting, but our previous analysis of normal modes
suggests the following modification of (\ref{5.46}) 
\ba
\la{5.47}
\Gamma_{\rm one-loop}^{\rm D3} &=  \underbrace{\Gamma_{\rm D}^{(2,0)}+2\Gamma_{\rm D}^{(2,1)}}_{AdS_{5}}
+\underbrace{\Gamma_{\rm N}^{(1,0)}
+2\Gamma_{\rm N}^{(1,\alpha_{2})}+2\Gamma_{\rm D}^{(1,\alpha_{2})}}_{S^{5}}\lp
- 2\,[
\wh\Gamma_{\rm N}^{(\frac{3}{2},\frac{1}{2}+\alpha_{2})}+\wh\Gamma_{\rm D}^{(\frac{3}{2},-\frac{1}{2}-\alpha_{2})}
+ \wh\Gamma_{\rm D}^{(\frac{3}{2},\frac{1}{2}-\alpha_{2})}+ \wh\Gamma_{\rm N}^{(\frac{3}{2},-\frac{1}{2}+\alpha_{2})}\lp
\ \ \ +\wh\Gamma_{\rm D}^{(\frac{3}{2},\frac{1}{2})}+\wh\Gamma_{\rm N}^{(\frac{3}{2},\frac{1}{2})}
+\wh\Gamma_{\rm D}^{(\frac{3}{2},-\frac{1}{2})}+\wh \Gamma_{\rm N}^{(\frac{3}{2},-\frac{1}{2})}], 
\ea
where the fermionic contribution is a sum of terms of the form (\ref{5.43}) and (\ref{5.45}).
One readily checks that (\ref{5.47}) gives indeed (\ref{5.2})
using (\ref{5.20}) and (\ref{5.21}). In fact,
\ba
\Gamma_{\rm D}^{(2,0)} &= -\frac{q^{3}}{1-q^{2}}, \qquad
\Gamma_{\rm D}^{(2,1)} = -\frac{q^{2}(1+q^{2})}{2(1-q^{2})}, \qquad
\Gamma_{\rm N}^{(1,0)} = -\frac{q}{1-q^{2}},\\
\Gamma_{\rm N}^{(1,\alpha_{2})} &= -\frac{q}{2(1-q^{2})}(\eta+\eta^{-1}), \qquad
\Gamma_{\rm D}^{(1,\alpha_{2})} = -\frac{q^{2}}{2(1-q^{2})}(\eta+\eta^{-1}), \\
\wh\Gamma_{\rm N}^{(\frac{3}{2},\frac{1}{2}+\alpha_{2})} &= -\frac{q^{2}}{2(1-q^{2})}\,\eta,\qquad
\wh\Gamma_{\rm D}^{(\frac{3}{2},-\frac{1}{2}-\alpha_{2})} = -\frac{q^{2}}{2(1-q^{2})}\,\eta^{-1}, \\
\wh\Gamma_{\rm D}^{(\frac{3}{2},\frac{1}{2}-\alpha_{2})} &= -\frac{q^{3}}{2(1-q^{2})}\,\eta^{-1},\qquad
\wh\Gamma_{\rm N}^{(\frac{3}{2},-\frac{1}{2}+\alpha_{2})} = -\frac{q}{2(1-q^{2})}\,\eta,\\
\wh\Gamma_{\rm D}^{(\frac{3}{2},\frac{1}{2})} &+\wh\Gamma_{\rm N}^{(\frac{3}{2},\frac{1}{2})}
+\wh\Gamma_{\rm D}^{(\frac{3}{2},-\frac{1}{2})}+\wh \Gamma_{\rm N}^{(\frac{3}{2},-\frac{1}{2})} 
= -\frac{q(1+q)^{2}}{2(1-q^{2})},
\ea
and total is 
\ba
\Gamma^{\rm D3}_{\oneloop} &= -\frac{1}{1-q^{2}}[q^{3}+q^{2}(1+q^{2})+q+q(\eta+\eta^{-1})+q^{2}(\eta+\eta^{-1})\lp
-q^{2}\eta-q^{2}\eta^{-1}-q^{3}\eta^{-1}-q\eta-q(1+q)^{2}] = q^{2}-q\,\eta^{-1}.
\ea
The separate and total Casimir energies are 
\ba
E_{c,\rm D}^{(2,0)} &=-\tfrac{11}{24} , \qquad
E_{c, \rm D}^{(2,1)} =-\tfrac{7}{12} , \qquad
E_{c, \rm N}^{(1,0)} = \tfrac{1}{24},\\
E_{c, \rm N}^{(1,\alpha_{2})} &= \tfrac{1}{24}-\tfrac{1}{8}\alpha_{2}^{2}, \qquad
E_{c, \rm D}^{(1,\alpha_{2})} =-\tfrac{1}{12}-\tfrac{1}{8}\alpha_{2}^{2} , \\
{\wh E}_{c, \rm N}^{(\frac{3}{2},\frac{1}{2}+\alpha_{2})} &= -\tfrac{1}{24}-\tfrac{1}{8}\alpha_{2}-\tfrac{1}{16}\alpha_{2}^{2},\qquad
{\wh E}_{c, \rm D}^{(\frac{3}{2},-\frac{1}{2}-\alpha_{2})} = -\tfrac{1}{24}+\tfrac{1}{8}\alpha_{2}-\tfrac{1}{16}\alpha_{2}^{2}, \\
{\wh E}_{c, \rm D}^{(\frac{3}{2},\frac{1}{2}-\alpha_{2})} &= -\tfrac{11}{48}+\tfrac{1}{4}\alpha_{2}-\tfrac{1}{16}\alpha_{2}^{2}\qquad
{\wh E}_{c, \rm N}^{(\frac{3}{2},-\frac{1}{2}+\alpha_{2})} = \tfrac{1}{48}-\tfrac{1}{16}\alpha_{2}^{2}, \\
{\wh E}_{c, \rm D}^{(\frac{3}{2},\frac{1}{2})} &+{\wh E}_{c, \rm N}^{(\frac{3}{2},\frac{1}{2})}
+{\wh E}_{c, \rm D}^{(\frac{3}{2},-\frac{1}{2})}+{\wh E}_{c, \rm N}^{(\frac{3}{2},-\frac{1}{2})} 
= -\tfrac{7}{24},
\ea
and
\ba
E^{\rm D3}_{c, \oneloop} &= -\tfrac{11}{24}-2\times\tfrac{7}{12}+\tfrac{1}{24}
+2\times(\tfrac{1}{24}-\tfrac{1}{8}\alpha_{2}^{2})+2\times(-\tfrac{1}{12}-\tfrac{1}{8}\alpha_{2}^{2})\lp
-2\bigg[
 -\tfrac{1}{24}-\tfrac{1}{8}\alpha_{2}-\tfrac{1}{16}\alpha_{2}^{2}
  -\tfrac{1}{24}+\tfrac{1}{8}\alpha_{2}-\tfrac{1}{16}\alpha_{2}^{2}
  -\tfrac{11}{48}+\tfrac{1}{4}\alpha_{2}-\tfrac{1}{16}\alpha_{2}^{2}
  + \tfrac{1}{48}-\tfrac{1}{16}\alpha_{2}^{2}-\tfrac{7}{24}\bigg]\lp
  =-\tfrac{1}{2}-\tfrac{1}{2}\alpha_{2}.
\ea
Quite remarkably, in the presence of the D3 brane, the total value of the supersymmetric Casimir energy 
$E_{c}$ is not zero but provides instead the linear in $\beta$ term required in (\ref{5.2}). It is thus responsible for the 
the puzzling factor $1/(\eta q)$ in (\ref{1.2}) and (\ref{2.9}), first discovered in \cite{Imamura:2024lkw}.

\section*{Acknowledgements}

I thank Arkady Tseytlin for many valuable discussions and clarifications, and Alejandro Cabo-Bizet for useful comments.
Financial support from the INFN grant GAST is acknowledged.

\appendix

\section{Thermal one-loop scalar determinant in $AdS_{2}$}
\la{app:der}

Let us focus on the non-Casimir part of the thermal scalar determinant in $AdS_{2}$. 
The effective action before thermal quotient is \cite{Bergamin:2015vxa}.
\be
\log\det(-\nabla^{2}+m^{2}) = -\int_{s_{\rm UV}}^{\infty}\frac{ds}{s}\int d^{2}\xi\, \sqrt{g}\, K(\xi,\xi; s), \qquad \vol(AdS_{2}) = -2\pi,
\ee
where $K$ is the heat kernel.
Let $D(\xi,\xi')$ be the the geodesic distance. In Poincare' patch $\xi=(x,y)$ we have 
\be
ds^{2} = \frac{dx^{2}+dy^{2}}{y^{2}}, \qquad \sqrt g = \frac{1}{y^{2}}, \qquad \cosh D(\xi, \xi')= 1+\frac{(y-y')^{2}+(x-x')^{2}}{2yy'}.
\ee
The heat kernel for a massive scalar field is \cite{Camporesi:1990wm}
\be
K(\xi, \xi'; s) = K(D(\xi,\xi'); s) = \frac{\sqrt 2\, e^{-(m^{2}+\frac{1}{4})s}}{(4\pi s)^{3/2}}\int_{D(\xi,\xi')}^{\infty}\frac{du\, u\, e^{-\frac{u^{2}}{4s}}}
{\sqrt{\cosh u-\cosh D(\xi,\xi')}}.
\ee
In thermal $AdS_{2}$ we  follow  \cite{Giombi:2008vd} and consider the group action ( we don't consider an angular twist)
\be
\gamma\cdot \xi = (q^{-1}x, q^{-1}y).
\ee
According to the method of images, we need
\bea
\wt K(\xi,\xi; s) &= \sum_{n\in\mathbb Z}K(D(\xi,\gamma^{n}\cdot \xi); s), \\
\cosh D(\xi,\gamma^{-n}\cdot \xi) &= 1+\frac{(x^{2}+y^{2})(1-q^{n})^{2}}{2y^{2} q^{n}}
\eea
Let us introduce on the quotient $AdS_{2}/\mathbb Z$  polar coordinates on the fundamental domain (see for instance 
 \cite{Giombi:2008vd} for $AdS_{3}$ case)
\be
x = r \cos\phi, \qquad y = r \sin\phi, \qquad r\in[1,e^{\beta}], \quad \phi\in[0,\pi], \qquad \sqrt g = \frac{1}{r\sin^{2}\phi}
\ee
\be
\cosh D(r,\phi) = 1+\frac{(1-q^{n})^{2}}{2q^{n}}\frac{1}{\sin^{2}\phi}.
\ee
We need to evaluate
\ba
\int\frac{dr d\phi}{r\sin^{2}\phi}\int_{D(r, \phi)}^{\infty}\frac{du\, u\, e^{-\frac{u^{2}}{4s}}}
{\sqrt{\cosh u-\cosh D(r,\phi)}} =2 \int_{1}^{q^{-1}}\frac{dr}{r}\int_{0}^{\pi/2}\frac{d\phi}{\sin^{2}\phi}\int_{D(r, \phi)}^{\infty}\frac{du\, u\, e^{-\frac{u^{2}}{4s}}}
{\sqrt{\cosh u-\cosh D(r,\phi)}}
\ea
When $\phi\in (0,\pi/2)$ we have 
\bea
D & \ge \text{arccosh}\bigg[1+\frac{(1-q^{n})^{2}}{2q^{n}}\bigg] = n\beta, \\
\frac{d\phi}{\sin^{2}\phi} &= \frac{1}{\sqrt{2} q^{n/2}(-1+q^{-n})}\frac{\sinh D}{\sqrt{\cosh D-\cosh(n\beta)}}dD.
\eea
By exchanging order of integration we obtain
\ba
& \int_{n\beta}^{\infty} dD\frac{\sinh D}{\sqrt{\cosh D-\cosh(n\beta)}}\int_{D}^{\infty}du \frac{u\, e^{-\frac{u^{2}}{4s}}}
{\sqrt{\cosh u-\cosh D}} \lp
= \int_{n\beta}^{\infty} du\, u\, e^{-\frac{u^{2}}{4s}}\int_{n\beta}^{u}dD \frac{\sinh D}{\sqrt{\cosh D-\cosh(n\beta)}\sqrt{\cosh u-\cosh D}} \lp
= \pi \int_{n\beta}^{\infty} du u\, e^{-\frac{u^{2}}{4s}} = 2\pi s e^{-\frac{n^{2}\beta^{2}}{4s}}.
\ea
So, considering a fixed value of $n\neq 0$ (one then sums over $n$ treating separately $n=0$, see \cite{Giombi:2008vd})
\ba
\log & \det(-\nabla^{2}+m^{2})\bigg|_{n} = -2 \int_{1}^{q^{-1}}\frac{dr}{r}\int_{0}^{\infty}\frac{ds}{s}
\frac{\sqrt 2\, e^{-(m^{2}+\frac{1}{4})s}}{(4\pi s)^{3/2}} \frac{1}{\sqrt{2} q^{n/2}(-1+q^{-n})}2\pi s e^{-\frac{n^{2}\beta^{2}}{4s}} \lp
= -\frac{q^{n/2}e^{-n\beta\sqrt{\frac{1}{4}+m^{2}}}}{n(1-q^{n})} =-\frac{1}{n} \frac{q^{nh}}{1-q^{n}},\qquad \qquad
h = \frac{1}{2}+\sqrt{\frac{1}{4}+m^{2}},
\ea
which confirms (\ref{4.54}) in $\kappa=0$ case by a direct calculation.

\section{Casimir energy and $\beta\ll 1$ expansion of partition function}
\la{app:Casimir}

Given the single particle partition function $Z_{\rm s.p.}(\beta)$ \footnote{The present discussion  
holds for the standard Casimir energy and also for the 
supersymmetric Casimir energy if $Z_{\rm s.p.}$ is the single particle superconformal index.}
we can determine the Casimir contribution from (\ref{4.56})
\be
E_{c} = \frac{1}{2}\zeta_{E}(-1), \qquad \zeta_{E}(z) = \frac{1}{\Gamma(z)}\int_{0}^{\infty}d\beta \beta^{z-1}Z_{\rm s.p.}(\beta).
\ee
This is a Mellin transform with inverse formula
\be
\la{B.2}
Z_{\rm s.p.}(\beta) = \frac{1}{2\pi i }\int_{c-i\infty}^{c+i\infty}dz\, \beta^{-z}\, \Gamma(z) \zeta_{E}(z),  
\ee
where $c$ in the analyticity strip of $\zeta_{E}(z)$. 
As shown in \cite{Gibbons:2006ij} one can consider the full partition function 
\be
Z(\beta)  =\PE[Z_{\rm s.p.}] =  \exp\sum_{n=1}^{\infty}\frac{1}{n}Z_{\rm s.p.}(n \beta),
\ee
and  (\ref{B.2}) gives the following integral representation
\be
\log Z(\beta) = \sum_{n=1}^{\infty}\frac{1}{n}
\frac{1}{2\pi i }\int_{c-i\infty}^{c+i\infty}dz\, (n\beta)^{-z}\, \Gamma(z)\zeta_{E}(z) = 
\frac{1}{2\pi i }\int_{c-i\infty}^{c+i\infty}dz\, \beta^{-z}\, \zeta(z+1)\Gamma(z)\zeta_{E}(z).
\ee
From the pole at $z=-1$ we find that 
the Laurent expansion of $\log Z(\beta)$ at small $\beta$ has then a linear in $\beta$ term of the form 
\be
\la{B.5}
\log Z(\beta) = \text{singular terms} +\#\beta^{0}+E_{c}\beta+\cdots.
\ee
Actually, $E_{c}$ can be extracted in an easier way by noting that $Z_{\rm s.p.}(\beta)$ has also a Laurent expansion around
$\beta=0$ 
and the coefficient of the linear in $\beta$ term should be  $-2E_{c}$. In fact, one has then 
\be
\log Z(\beta) = \sum_{n=1}^{\infty}\frac{1}{n}Z_{\rm s.p.}(n \beta) = \cdots+\zeta(0) (-2E_{c})\beta+\cdots,
\ee
in agreement with (\ref{B.5}). We have thus the simple relation 
\be
E_{c} = -\frac{1}{2} Z_{\rm s.p.}(\beta)\bigg|_{\beta\ \text{term}},
\ee
which is quite efficient for calculations. For instance, for a scalar field in $AdS_{2}$
\be
Z_{\rm s.p.}(\beta) = \frac{e^{-\beta h}}{1-e^{-\beta}} = \frac{1}{\beta}+(\tfrac{1}{2}-h)+\tfrac{1}{12}(1-6h+6h^{2})\beta+\cdots,
\ee
consistently with (\ref{4.59}). As another example, for a conformally coupled scalar on $S^{1}\times S^{3}$ one has 
\be
Z_{\rm s.p.}(\beta) = \frac{e^{-\beta}(1-e^{-2\beta})}{(1-e^{-\beta})^{4}} = \frac{2}{\beta^{3}}-\frac{1}{120}\beta+\cdots,
\ee
reproducing the known value $E_{c}=\frac{1}{240}$.

\section{$OSp(4^{*}|4)$ ultra-short multiplet in the presence of the D3 brane}
\la{app:osp}

Unitary representations of $OSp(2n^{*}|2m)$ were discussed in \cite{Gunaydin:1990ag} using an oscillator representation.
The specific case $OSp(4^{*}|4)$ was summarized in  \cite{Faraggi:2011bb}. Here we present in full details the case of its 
ultra-short representation and its reduction to the smaller representation in Table \ref{tab:2} in the presence of the 
 D3 giant.
 
\subsection{Structure of $OSp(4^{*}|4)$ and oscillator representation}

The group $OSp(4^{*}|4)\supset SO(4^{*})\times USp(4)$  has a Jordan structure with respect the the maximal subgroup 
$G^{0}=U(2|2)\supset U(2)\times U(2)$. At Lie superalgebra level
\be
\mk{osp}(4^{*}|4) = \mk{g}^{-1}\oplus \mk{g}^{0}\oplus \mk{g}^{+1}, 
\qquad [\mk{g}^{a}, \mk{g}^{b}]\subseteq \mk{g}^{a+b}, \qquad \mk{g}^{a}=0\ \text{for } |a|>1.
\ee
Generators are split into 
\be
\la{C.2}
M_{AB}\in \mk{g}^{-1}, \qquad M\indices{^{A}_{B}}\in \mk{g}^{0}, \qquad M^{AB}\in \mk{g}^{+1},
\ee
where the index $A=(i, \mu)$ with $i=1,2$, $\mu=1,2$ is in the fundamental of $U(2|2)$ and we assign $\deg i= 0$, $\deg \mu=1$.
The Lie superalgebra is realized by introducing a pair of super-oscillators \footnote{The general case requires several copies of these two 
super-oscillators \cite{Gunaydin:1990ag}, but for the discussion of the ultra-short representations a single pair is enough.} 
\be
\xi_{A} = \binom{a_{i}}{\alpha_{\mu}}, \qquad \eta_{A} = \binom{b_{i}}{\beta_{\mu}},
\ee
where the non-vanishing (anti) commutators are 
\be
[a_{i}, a^{j}] =[b_{i}, b^{j}]  =  \delta_{i}^{j}, \qquad 
\{\alpha_{\mu}, \alpha^{\nu}\} = \{\beta_{\mu}, \beta^{\nu}\} = \delta_{\mu}^{\nu}.
\ee
Generators in (\ref{C.2}) can be written 
\ba
M_{AB} &= \xi_{A}\eta_{B}-\eta_{A} \xi_{B}, \quad
M^{AB} = \eta^{B}\xi^{A}-\xi^{B} \eta^{A}, \quad
M\indices{^{A}_{B}} = \xi^{A}\xi_{B}+(-1)^{\deg A\deg B}\eta_{B} \eta^{A}.
\ea
The bosonic bilinears 
\be
M_{ij} = a_{i}b_{j}-a_{j}b_{i}, \quad M^{ij}=a^{i}b^{j}-a^{j}b^{i}, \quad M\indices{^{i}_{j}} = a^{i}a_{j}+b_{j}b^{i},
\ee
generate (at group level) $SO(4^{*}) = SL(2,\mathbb R)\times SU(2)$. In particular, 
\be
B^{-} = M_{12}, \qquad B^{+} = M^{12}, \qquad 
B^{0}=\frac{1}{2}M\indices{^{i}_{i}} = \frac{1}{2}(a^{i}a_{i}+b_{i}b^{i}),
\ee
are generators of $SL(2,\mathbb R)$ obeying 
\be
[B^{+},B^{-}] = 2B^{0}, \qquad [B^{0}, B^{\pm}] = \pm  B^{\pm}.
\ee
The bilinears
\be
I\indices{^{i}_{j}} = M\indices{^{i}_{j}}-\frac{1}{2}\delta^{i}_{j}M\indices{^{k}_{k}},
\ee
generate  $SU(2)$. The trace condition is $I\indices{^{2}_{2}} = -I\indices{^{1}_{1}}$ and one has standard algebra
\be
[I\indices{^{1}_{1}}, I\indices{^{1}_{2}}] = I\indices{^{1}_{2}}, \qquad
[I\indices{^{1}_{1}}, I\indices{^{2}_{1}}] = -I\indices{^{2}_{1}}, \qquad
[I\indices{^{1}_{2}}, I\indices{^{2}_{1}}] = 2I\indices{^{1}_{1}}.
\ee
Bilinears in the fermionic oscillators are \footnote{We use a different font to make it clear whether a certain operator with explicit indices has them  of Latin or Greek type.}
\be
\la{C.11}
\sfM_{\mu\nu} = \alpha_{\mu}\beta_{\nu}-\beta_{\mu}\alpha_{\nu}, \qquad
\sfM^{\mu\nu} = \beta^{\nu}\alpha^{\mu}-\alpha^{\nu}\beta^{\mu}, \qquad
\sfM\indices{^{\mu}_{\nu}} = \alpha^{\mu}\alpha_{\nu}-\beta_{\nu}\beta^{\mu}.
\ee
They are $3+3+4=10$ \footnote{Oscillators are fermionic so for instance $\sfM_{11}\neq 0$.}
and generate $USp(4) = SO(5)_{R}$.
Fermionic bilinears are the following $4\times 4 = 16$ (real) supercharges
\be
Q_{i\mu} = a_{i}\beta_{\mu}-b_{i}\alpha_{\mu}, \qquad
Q^{i\mu} = a^{i}\beta^{\mu}-b^{i}\alpha^{\mu}, \qquad
S\indices{^{i}_{\mu}} = a^{i}\alpha_{\mu}+b^{i}\beta_{\mu},\qquad
S\indices{^{\mu}_{i}} = a_{i}\alpha^{\mu}+b_{i}\beta^{\mu}.
\ee

\subsection{The ultra short multiplet}

The shortened multiplet in (\ref{3.7}) contains only spin $0$ and spin $\frac{1}{2}$ states. It is built
starting from the oscillator vacuum $\vac$ annihilated by $a_{i}, b_{i}, \alpha_{\mu}, \beta_{\mu}$. With respect to 
$SL(2, \mathbb R)\times SU(2)$ it obeys
\be
B^{0}\vac = \vac, \qquad I\indices{^{i}_{j}}\vac = 0 .
\ee
Acting on it with the generators of $SO(5)$ in (\ref{C.11}) we get the additional $3+1$ states
\be
(\alpha^{\mu}\beta^{\nu}+\alpha^{\nu}\beta^{\mu})\vac, \qquad \alpha^{2}\alpha^{1}\beta^{2}\beta^{1}\vac,
\ee
that together with $|0\rangle$ give the $\bm 5$ of $SO(5)_{R}$. Thus we have a first lowest weight states in the  $SL(2,\mathbb R)\times SU(2)\times SO(5)$ 
representation, \cf (\ref{5.38}),
\be
\vp (1, \bm{1}, \bm{5}).
\ee
Acting with the supercharges $Q^{i\mu}$ on $\vac$ we get 4  non vanishing states with dilatation eigenvalue $h=\frac{3}{2}$. 
Taking their orbit with $SU(2)$ and $SO(5)_{R}$ generators gives 8 states. These have $(J\indices{^{1}_{2}})^{2}$ 
and definite $J\indices{^{1}_{1}}$ eigenvalue equal to $\pm \frac{1}{2}$ and belong to the 
 $\bm{2}$ of $SU(2)$. Let us list their explicit form 
\ba
\la{C.16}
\def\arraystretch{1.3}
\begin{array}{rr}
 & J\indices{^{1}_{1}} \\
(-\alpha^1 b^1+\beta^1 a^1)\,\vac   & +\frac{1}{2} \\
(-\alpha ^1b^2+\beta^1 a^2)\,\vac  & +\frac{1}{2} \\
(-\alpha ^2 b^1 +\beta ^2 a^1)\,\vac & -\frac{1}{2} \\
(-\alpha ^2 b^2+\beta^2 a^2)\,\vac & -\frac{1}{2} \\
\midrule
(\alpha ^1\beta ^2\beta ^1a^1+\alpha ^2\alpha^1\beta ^1 b^1)\,\vac & +\frac{1}{2}\\
(\alpha ^1\beta ^2\beta ^1a^2+\alpha^2\alpha ^1\beta ^1b^2 )\,\vac & -\frac{1}{2} \\
(\alpha ^2\alpha ^1\beta ^2b^1+\alpha ^2\beta ^2\beta ^1a^1 )\,\vac & +\frac{1}{2} \\
(\alpha ^2\alpha^1\beta ^2b^2+\alpha ^2\beta ^2\beta ^1a^2 )\,\vac & -\frac{1}{2}
\end{array}
\ea
These 8 states correspond to the lowest weight states, \cf (\ref{5.38}), 
\be
\psi (\tfrac{3}{2}, \bm{2}, \bm{4}).
\ee
Next, we consider the states
\be
\la{C.18}
Q^{i\mu}Q^{j\nu}\vac.
\ee
These have $h=2$. Some of them are conformal descendant of $\vp(1, \bm{1}, \bm{5})$. Additional primary states are the three ones
\ba
\def\arraystretch{1.3}
\begin{array}{rr}
& J\indices{^{1}_{1}} \\
(\alpha ^1\beta ^2a^1b^1+\alpha ^2\alpha ^1b^1b^1-\alpha ^2\beta ^1a^1b^1
+\beta ^2\beta ^1a^1a^1)\,\vac & 1 \\
(\alpha ^1\beta ^2a^1b^2+\alpha ^2\alpha ^1b^2b^1-\alpha ^2\beta ^1a^2b^1 +
\beta ^2\beta ^1a^2a^1 )\,\vac & 0\\
(\alpha ^1\beta ^2a^2b^2+\alpha ^2\alpha ^1b^2b^2-\alpha ^2\beta ^1a^2b^2
+\beta ^2\beta^1a^2a^2 )\, \vac & -1
\end{array}
\ea
and build the last lowest weight representation in (\ref{5.38}) (one can check that $A^{\mu\nu}$ give zero) 
\be
\la{C.20}
\phi (2, \bm{3}, \bm{1}).
\ee

\subsection{Breaking R-symmetry $SO(5)_{R}\to SO(2)_{R}\times SO(3)_{R}$}

The $SO(5)_{R}$ generators
\ba
\sfL^{1} &= -\frac{i}{2}(\sfM\indices{^{1}_{2}}-\sfM\indices{^{2}_{1}}),\qquad
\sfL^{2} = -\frac{1}{2}(\sfM\indices{^{2}_{1}}+\sfM\indices{^{1}_{2}}),\qquad
\sfL^{3} = \frac{1}{2}(\sfM\indices{^{1}_{1}}- \sfM\indices{^{2}_{2}}),
\ea
obey
\be
[\sfL^{i},\sfL^{j}] = i\eps^{ijk}\,\sfL^{k},
\ee
and generate $SO(3)_{R}$. Also
\be
[\sfL^{i}, \sfR]=0, \qquad \sfR =\frac{1}{2}( \sfM\indices{^{1}_{1}}+ \sfM\indices{^{2}_{2}}),
\ee
and $\sfR$ generates $SO(2)_{R}$ (commuting with $SO(3)_{R}$).

The state $\vac$ is part of $\bm 5$ of $SO(5)_{R}$ and is a singlet of $SO(3)_{R}$ with $SO(2)_{R}$ quantum number $\sfR = -1$.
It is the scalar $\vp_{-}$ in Table \ref{tab:2}. Same for the state $\alpha^{2}\alpha^{1}\beta^{2}\beta^{1}\vac$ but 
with $SO(2)_{R}$ number $\sfR = +1$. It is the scalar $\vp_{-}$ in Table \ref{tab:2}.
The three states $(\alpha^{\mu}\beta^{\nu}+\alpha^{\nu}\beta^{\mu})\vac$ are in the $\bm{3}$ of $SO(3)$ and 
have $SO(2)_{R}$ charge $\sfR=0$. We denote representations of states in the presence of the D3 brane by four labels
\be
(B^{0}\equiv h, \dim SU(2), \sfR, \dim SO(3)_{R}).
\ee
The five scalars corresponding to $S^{5}$ fluctuations are then 
\be
\la{C.25}
\begin{array}{rl}
\vac & \vp_{-}(1,\bm{1},-1,\bm{1}) \\
(\alpha^{\mu}\beta^{\nu}+\alpha^{\nu}\beta^{\mu})\vac &  \vp_{0}(1,\bm{1},0,\bm{3})  \\
\alpha^{2}\alpha^{1}\beta^{2}\beta^{1} \vac &  \vp_{+}(1,\bm{1},+1,\bm{1}) 
\end{array}
\ee
Let us look at the 16 supercharges. We have conformal weights
\bea
& [B^{0},Q^{i\mu}] = \frac{1}{2}Q^{i\mu}, \qquad  [B^{0},Q_{i\mu}] =-\frac{1}{2}Q_{i\mu}, \\
& [B^{0},S\indices{^{i}_{\mu}}] = \frac{1}{2}S\indices{^{i}_{\mu}}, \qquad  [B^{0},S\indices{_{i}^{\mu}}] = -\frac{1}{2}S\indices{_{i}^{\mu}},
\eea
and $\sfR$ charges
\bea
& [\sfR,Q^{i\mu}] = \frac{1}{2}Q^{i\mu}, \qquad  [\sfR,Q_{i\mu}] =-\frac{1}{2}Q_{i\mu}, \\
& [\sfR,S\indices{^{i}_{\mu}}] = -\frac{1}{2}S\indices{^{i}_{\mu}}, \qquad  [\sfR,S\indices{_{i}^{\mu}}] = \frac{1}{2}S\indices{_{i}^{\mu}}.
\eea
The supercharges that are preserved in the presence of the wrapped D3 brane are those with $B^{0}=\sfR$ \cite{Imamura:2021ytr} \ie 
$Q^{i\mu}$ and $Q_{i\mu}$. Starting from $\vac$ we consider $Q^{i\mu}\vac$ and get four fermionic states with 
$SO(2)_{R}\times SO(3)_{R}$ quantum numbers
\ba
\def\arraystretch{1.3}
\begin{array}{lll}
 & \sfR & \sfL^{3} \\
(-\alpha^1 b^1+\beta^1 a^1)\,\vac   & -\frac{1}{2} & +\frac{1}{2}\\
(-\alpha ^1b^2+\beta^1 a^2)\,\vac  & -\frac{1}{2} & -\frac{1}{2}\\
(-\alpha ^2 b^1 +\beta ^2 a^1)\,\vac & -\frac{1}{2} & +\frac{1}{2}\\
(-\alpha ^2 b^2+\beta^2 a^2)\,\vac  & -\frac{1}{2} &-\frac{1}{2} 
\end{array}
\ea
This are the four states in second line of Table \ref{tab:2}
\be
\psi_{-}(\tfrac{3}{2}, \bm{2}, -\tfrac{1}{2}, \bm{2}).
\ee
and of course are the first four in (\ref{C.16}) corresponding to $\psi\to (\psi_{-}, \psi_{+})$ under R-symmetry breaking.

The same relation applies to descendants. Consider for instance the application of $B^{+}$ to the scalar $\vp_{-}$. Since $B^{+}$ commutes with 
generators of $SU(2)$ and also with $\sfR$ and generators of $SO(3)$ we have 
\be
B^{+}(1, \bm{1},-1, \bm{1}) = (2,\bm{1},-1, \bm{1}),
\ee 
which is the next mode of $\vp_{-}$. We also have 
\be
[B^{+}, Q^{i\mu}] = 0.
\ee
Hence, supersymmetry relates the $n$-th mode of $\vp_{-}$, with $h=1+n$, to the $n$-th mode of $\psi_{-}$, with $h=\frac{3}{2}+n$.

Let us now apply two supersymmetries to $\vp_{-}$. We get the states
\be
Q^{i\mu}Q^{j\nu}\vac.
\ee
As in the discussion after (\ref{C.18}), we have 6 states. Three are level 1 conformal  descendants  $B^{+}\vp_{0}$ and three are the states in $(2,\bm{3},\bm{1})$ in (\ref{C.20}).
Two examples illustrating the two cases are 
\bea
& B^{+}\alpha^{1}\beta^{1}\vac = -Q^{2,1}Q^{1,1}\vac, \\ 
& (\alpha ^1\beta ^2a^1b^1+\alpha ^2\alpha ^1b^1b^1-\alpha ^2\beta ^1a^1b^1
+\beta ^2\beta ^1a^1a^1)\,\vac = Q^{1,2}Q^{1,1}\vac.
\eea
At level 3, we find 4 independent states of the form 
\be
Q^{i\mu}Q^{j\nu}Q^{k\rho}\vac,
\ee
and they are (linear combinations of ) the $B^{+}$ descendents of the last four states in (\ref{C.16}), \ie of the spinor $\psi_{+}$.
Finally, there is a single state of the form 
\be
Q^{i\mu}Q^{j\nu}Q^{k\rho}Q^{\ell\l}\vac.
\ee
It is obtained for example with this product of 4 supercharges
\be
Q^{1,1}Q^{2,2}Q^{1,2}Q^{2,1}\vac = \alpha^{2}\alpha^{1}\beta^{2}\beta^{1}\bigg(
(a^{1})^{2}(b_{2})^{2}
-2a^{2}a^{1}b^{2}b^{1}
+(a^{2})^{2}(b^{1})^{2}
\bigg)\vac = (B^{+})^{2}\alpha^{2}\alpha^{1}\beta^{2}\beta^{1}\vac,
\ee
and is the  level 2 conformal descendent of $\vp_{+}$, \cf last line in  (\ref{C.25}). Our analysis is summarized in (\ref{5.40}).

\bibliography{BT-Biblio}
\bibliographystyle{JHEP-v2.9}
\end{document}